\documentclass[11pt]{article}
\usepackage{msh}
\usepackage{graphicx}
\usepackage{dcolumn}
\usepackage{bm}
\usepackage{amsmath,amssymb}
\usepackage[version=4]{mhchem}
\usepackage{siunitx}
\usepackage{longtable,tabularx}
\usepackage{float}
\usepackage{subcaption}
\usepackage{siunitx}
\usepackage[short]{optidef}
\usepackage{dsfont}
\usepackage{newtxtext}
\usepackage{newtxmath}
\usepackage{natbib}
\usepackage{hyperref}
\usepackage{makecell}
\usepackage{color}

\title{\Large\sffamily\textbf{ Data-driven nonlinear aerodynamics models with certifiably optimal boundedness properties }}

\author{A. Leonid Heide,$^1$
Shih-Chi Liao,$^2$
Sergio Castiblanco-Ballesteros,$^3$
Gustaaf B Jacobs,$^3$
Peter Seiler,$^2$
\and Maziar S. Hemati$^1$}

\date{$^1$Department of Aerospace Engineering and Mechanics, University of Minnesota\\
$^2$Department of Electrical and Computer Engineering, University of Michigan, Ann Arbor\\
$^3$Department of Aerospace Engineering, San Diego State University}

\begin{document}

\maketitle
\begin{abstract} 
Obtaining predictive low-order models is a central challenge in fluid dynamics. Data-driven frameworks have been widely used to obtain low-order models of aerodynamic systems; yet, resulting models tend to yield predictions that grow unbounded with time. Recently introduced stability-promoting methods can facilitate the identification of bounded models, but tend to require extensive brute-force tuning even in the context of simple academic systems. Here, we show how recent theoretical advances in the long-term boundedness of dynamical systems can be integrated into data-driven modeling frameworks to ensure that resulting models will yield bounded predictions of incompressible flows. Specifically, we propose to solve a specific set of convex semi-definite programming problems to (i)~certify whether a system admits a globally attracting bounded set for the chosen modeling parameters, and (ii) compute a model with the optimal (tightest) bound on this globally attracting set. We demonstrate the approach via integration within the sparse identification of nonlinear dynamics~(SINDy) modeling framework. Application on two low-order benchmark problems establishes the merits of the approach. We then apply our approach to obtain a low-order (6-mode) model of unsteady separation over a NACA-65(1)-412 airfoil at $Re=20,0000$---a flow that has been notoriously difficult to model using data-driven methods. The resulting model is found to accurately predict the dynamics of unsteady separation, with model predictions remaining bounded indefinitely. We anticipate this work will benefit future efforts in modeling strongly nonlinear flows, especially in settings where physically-viable long-term forecasts are paramount.
\end{abstract}

\section{Introduction}
Low-order models are an enabling tool in fluid dynamics, providing unique opportunities for rapid prediction and forecasting, estimation and control, and design optimization~\citep{bennerSurvey,Brunton2015,Rowley2017}.
However, most use-cases require that low-order models  remain faithful to the physics and underlying dynamics of the flows under consideration.
This can be especially challenging in aerodynamics applications because the difficulty of crafting a low-order model that faithfully captures the essential physical and dynamical processes of a flow grows with increasing Reynolds number~\citep{Moehlis2004,cavalieri2022}.

In fluid dynamics, low-order models have traditionally been obtained within the framework of projection-based model reduction~\citep{holmesBook}, whereby the Navier-Stokes equations are projected onto a low-dimensional subspace (e.g., Galerkin projection onto a subspace spanned by dominant proper orthogonal decomposition (POD) modes).
Indeed, projection-based reduced-order models~(ROMs) have proven a powerful tool for distilling the complex dynamics of unsteady flows into a compact set of ordinary differential equations that support real-time analysis, optimization, and predictive control~\citep{Rowley2017,Taira2017,Brunton2020,Taira2020}.

Despite demonstrated successes in obtaining simplified yet physically faithful models in fluid dynamics, projection-based model reduction can be difficult to implement in practice.
Projection-based approaches often demand full access to the governing equations and their discretizations---an intrusive requirement that limits their applicability, especially when working with complex computational fluid dynamics~(CFD) codes~\citep{Peherstorfer2016}.
Projection-based model reduction can also result in models that diverge due to neglected energy transfers~\citep{Cazemier1998,Schlegel2015}.

In contrast, data-driven frameworks---such as sparse identification of nonlinear dynamics~(SINDy) \citep{Brunton_SINDy,Loiseau2018}, physics-informed neural networks~(PINNs)\citep{Raissi2019,Thuerey2020}, operator learning~\citep{Li2021DeepONet}, and operator inference~\citep{Peherstorfer2016,kramerARFM2024}---are non-intrusive, inferring compact dynamical systems directly from data generated by CFD or measured in experiments.
The data-driven approach enables rapid model construction and deployment even when a first-principles description is unavailable or only partially observed~\citep{Brunton_SINDy,Raissi2019}. 

Nonetheless, low-order models obtained from data-driven frameworks have their own limitations as well, often failing to generalize outside training datasets and commonly yielding unphysical energy growth under extrapolation~\citep{Duraisamy2019,Karniadakis2021}.
As such, much work has focused on ensuring that data-driven models adhere to known physical laws (e.g.,~conservation of mass, momentum, and energy).
In the case of incompressible flows, the quadratic nature of the convective nonlinearity in the Navier-Stokes equations admits energy transfer between modal triads---so-called \emph{triadic interactions}~\citep{schmidt}.
Further, the divergence-free condition on the velocity field along with mild assumptions regarding boundary conditions results in the convective nonlinearity being \emph{lossless}---i.e.,~the quadratic terms are energy-preserving, only serving to redistribute kinetic energy without net production or dissipation. 
Data-driven methods have been developed to impose these physics-based constraints within modeling frameworks~\citep{Loiseau2018}; however, the enforcement of lossless quadratic terms is not sufficient to guarantee that resulting models will maintain physically viable predictions over long-time horizons.
Indeed, it is common for such models to yield predictions that grow unbounded with time---an indicator of unphysical behavior.
This is particularly problematic in flow control and estimation applications: a ROM that is not globally bounded may diverge or exit its regime of validity, undermining any controller or estimator based on it.

Notions of stability and boundedness have been investigated and developed to contend with the possibility of this unphysical feature in the \emph{dynamics} of low-order models.
Methods have been proposed for efficiently analyzing the asymptotic stability of quadratic ROMs~\citep{kramerSIAM2021,enayati2025} or for directly identifying asymptotically stable ROMs from data~\citep{goyal2024}.
However, asymptotic stability is often a stronger condition than would be desired in many aerodynamics applications where trajectories remain bounded with time, but do not necessarily asymptote to an equilibrium point (e.g.,~wake shedding from an airfoil).
For such systems, it is natural to consider notions of long-time boundedness of trajectories~\citep{khalil2002nonlinear}.

Prior work by~\cite{Schlegel2015} studied the problem of long-time boundedness and established that quadratic dynamical systems whose quadratic nonlinearity is lossless can admit a \emph{trapping region}---a monotonically attracting invariant set within which all trajectories remain indefinitely.
The Schlegel and Noack Trapping Theorem provides conditions for the existence of a trapping region, which can be used to prove the long-time boundedness of projection-based ROMs for incompressible flows via non-convex optimization~\citep{Schlegel2015}.
The theorem also provides a (conservative) upper-bound on the size of the trapping region, if one exists.
Recent work by~\cite{Kaptanoglu2021} has leveraged the Schlegel and Noack Trapping Theorem within the SINDy modeling framework to facilitate the identification of bounded models from data.
This \textit{Trapping SINDy} approach resorts to using gradient descent and relax-and-split methods~\citep{Champion2019} for numerical solution of the modeling problem.
The authors have provided a Python implementation of their method, which includes several benchmark examples that demonstrate the method's efficacy in yielding bounded models from data~\citep{desilva2020, Kaptanoglu2022}.
Recent work by~\cite{Peng} have extended the Trapping SINDy framework to consider quadratic systems whose nonlinearity weakly violates the lossless constraint, using local notions of long-time boundedness to enable modeling of a larger class of systems.

In our recent work~\citep{liao2024}, we have made theoretical advances that improve upon the Schlegel and Noack Trapping Theorem and establish a pair of convex semi-definite programming~(SDP) conditions that have been shown to yield trapping regions that are significantly smaller than those from earlier analytical estimates.
The first SDP poses a feasibility problem to certify \emph{whether} a monotonically attracting trapping region exists. If this problem is infeasible, one can formally conclude that no trapping region is admissible for the given system. 
When the feasibility condition is satisfied, a second (dual) SDP is solved to \emph{minimize} the radius of the invariant ball, thereby \emph{proving} that the dynamical system is confined by the smallest possible Euclidean sphere. 
Since there is a unique optimal solution for the radius of the smallest invariant ball---by virtue of convexity---it follows that this optimal trapping region size can provide a distinct dynamical boundedness property that should be captured by a low-order model in order to be \emph{dynamically consistent} with the governing equations, in as much as any physical property should be captured for a model to be \emph{physically consistent} with the governing equations.

In this paper, we show how the Trapping-SDP conditions from~\cite{liao2024} can be integrated within data-driven modeling frameworks to formally guarantee long-time boundedness in the modeling process, and to (at least approximately) achieve dynamical consistency as needed to produce physically-viable predictions over long-time horizons.
Our proposed Trapping-SDP modeling framework consists of an alternating (block coordinate descent) minimization that cycles between the solution of two convex optimization problems---(i)~a regression problem, and (ii)~a stability-enforcement problem---both of which can be solved efficiently using available convex optimization software packages.

To focus the discussion, we demonstrate this integration within the SINDy modeling framework---though integration within alternative data-driven modeling frameworks follows similarly.
The proposed Trapping-SDP modeling approach is first presented on two canonical examples that allow direct comparison with known governing equations and associated optimal trapping regions.
We then apply our approach to obtain a 6-mode ROM of unsteady flow separation over a NACA 65(1)-412 airfoil at a Reynolds number of $Re=20,000$ and angle-of-attack $\alpha=4^\circ$ based on data from direct numerical simulations.
In each scenario, we obtain models with guaranteed long-time boundedness properties and report the optimal trapping regions for the given set of parameters.
All identified models are found to be predictive for short- and long-time horizons when applied to off-training validation tests.

The paper is organized as follows: In Section~\ref{sect_losslessquadROMS}, we review requisite background on ROMs for incompressible flows, long-time boundedness, trapping regions, trapping SDPs, and SINDy.
In section \ref{sec:methodMain}, we present our Trapping-SDP modeling framework within the context of SINDy.
Section~\ref{sec:numericalEx} reports on the application of the Trapping‐SDP modeling framework on two numerical benchmark systems. 
Application to data-driven model reduction of separated flow is reported in Section~\ref{sec:AirfoilResults}.
Conclusions are summarized in Section~\ref{sec:conclusion}.

\section{Background and Preliminaries}
\label{sect_losslessquadROMS}
\subsection{Reduced-Order Models of Incompressible Flows}
Before presenting our method, we first establish the essential system properties of reduced-order models (ROMs) for incompressible flows. 
Reduced-order modeling aims to approximate the (infinite-dimensional) partial differential equations (PDEs) that govern the system with a low (finite) dimensional representation. A common model reduction technique is the Galerkin projection, i.e., an orthogonal projection of these PDEs onto a finite-dimensional subspace spanned by a set of basis functions. 
In incompressible flows, the nonlinear (convective) term of the Navier--Stokes equations merely redistributes energy among the flow's degrees of freedom---a so-called \emph{lossless} property of the convective nonlinearity.
Galerkin projection of the incompressible Navier-Stokes equations onto an $n$-dimensional subspace yields ordinary differential equations (ODEs) of the form
\begin{equation} \label{eq:Dynamic_System}
    \dot{x}(t) = c + Lx(t) + f(x(t)),
\end{equation}
where $x(t) \in \mathbb{R}^n$ is the reduced state, the function, $ f: \mathbb{R}^{n} \to \mathbb{R}^{n} $ is quadratic, with its $i$-th entry given by: $f_i(x) = x^\top Q^{(i)} x \ ( i = 1, \ldots, n)$,  the matrix $L \in \mathbb{R}^{n \times n} $ contains the linear terms, and $ c \in \mathbb{R}^n $ is a constant vector. The quadratic term matrices  $ \{ Q^{(i)} \}_{i=1}^n \in \mathbb{R}^{n \times n}$ here are symmetric.
If the kinetic energy of the system \eqref{eq:Dynamic_System} is defined as
\begin{equation}\label{eq:Dynamic_SystemEnergy}
    K_0(x) = \tfrac{1}{2}\|x\|^2,
\end{equation}
then the nonlinearity $f(x(t))$ is said to be \emph{lossless} if
\begin{equation}  \label{eq:LosslessQuad}
    x^\top f(x) = 0, \quad \forall x \in \mathbb{R}^n.
\end{equation}
This indicates that $f$ neither generates nor dissipates energy, but instead redistributes it among the components of $x$. Ensuring energy conservation requires that the quadratic term matrices $Q^{(i)}$ comprising the nonlinear term $f(x(t))$ satisfy the condition
\begin{equation} \label{eq:losslessCond}
     \sum_{i,j,k=1}^n Q_{jk}^{(i)} x_i x_j x_k = 0, \quad \forall x \in \mathbb{R}^n.
\end{equation}
Under the lossless condition, the time derivative of the energy \eqref{eq:Dynamic_SystemEnergy} along trajectories of the system \eqref{eq:Dynamic_System} simplifies to
\begin{equation} \label{eq:Energy_LosslessSystem}
    \frac{d}{dt}K_0(x(t)) = x(t)^\top \left(c + L_Sx(t)\right),
\end{equation}
where $ L_S = \frac{1}{2}(L + L^\top) $ denotes the symmetric part of $ L $. Equation \eqref{eq:Energy_LosslessSystem} indicates that the quadratic nonlinearity $f$ does not affect the energy evolution.

If the system's coordinates are translated by a constant vector $ m \in \mathbb{R}^n $, where $ y = x - m $, the system dynamics \eqref{eq:Dynamic_System} become
\begin{equation} \label{eq:sys_shifted}
    \dot{y}(t) = d(m) + A(m) y(t) + f(y(t)),
\end{equation}
where $ d(m) = c + Lm + f(m) $, and the matrix $ A(m) $ is given by
\begin{equation}
\label{eq:A_m}
    A(m) = L + 2\begin{pmatrix} m^\top Q^{(1)} \\ \vdots \\ m^\top Q^{(n)} \end{pmatrix}.
\end{equation}
The quadratic nonlinearity $ f(y) $ in the shifted system \eqref{eq:sys_shifted} remains identical to that in the original system \eqref{eq:Dynamic_System}, ensuring that the lossless property is preserved after translation. 
The symmetric part of $ A(m) $, denoted $ A_S(m) $, is defined as
\begin{equation} \label{eq:As}
    A_S(m) =\frac{1}{2} (A(m) + A(m)^\top)=  L_S - \sum_{k=1}^n m_k Q^{(k)},
\end{equation}
where $ m_k $ is the $ k $-th element of the translation vector $ m $. 
The energy in the shifted coordinates is $ K_{m}(y) = \frac{1}{2} \| y \|^2 $, and its time derivative is
\begin{equation} \label{eq:dKmdt}
    \frac{d}{dt} K_{m}(y(t)) = y(t)^\top d(m) + y(t)^\top A_s(m) y(t).
\end{equation}
Further details on Galerkin projection and the derivation of lossless ROMs can be found in appendix~\ref{apdx:triadic} and \citep{Loiseau2018,Schlegel2015}.

\subsection{Long-Time Boundedness and Trapping Regions} \label{sect:SystemBoundedness}
In reduced-order modeling of fluid dynamics, ensuring solutions remain bounded is a crucial problem, as unbounded behaviors correspond to nonphysical or unrealistic phenomena. Here, we focus on \emph{globally uniformly ultimately bounded} systems~\citep{khalil2002nonlinear}, which means there exists a region such that trajectories from any initial condition eventually lie within a ball of radius $\beta > 0$. Formally, for a nonlinear system \eqref{eq:Dynamic_System}, boundedness  the following holds for some $\beta>0$:
\begin{equation*}
    \|{x}(t)\| \leq \beta 
    \quad \forall\, t \geq t_0 + T\bigl(\|{x}_0\|\bigr),
\end{equation*}
where $T(\cdot)$ is a function depending on the initial condition ${x}_0$ but not on the initial time $t_0$.

A practical way to verify boundedness in lossless quadratic systems is through the concept of a \emph{trapping region}~\citep{Schlegel2015}, a compact, forward-invariant set $D\subseteq\mathbb{R}^n$ that, once entered by a trajectory, cannot be exited. If the energy is strictly decreasing outside $D$ (i.e. $\frac{dK}{dt}<0$), the trapping region is said to be \emph{globally monotonically attracting}, and hence all trajectories eventually enter $D$. 

Lorenz ~\citep{lorenz1963deterministic} provided a condition for the existence of such a trapping region for lossless systems like~\eqref{eq:Dynamic_System}, stating that if the symmetric part of the linear operator satisfies $ L_S = \frac{1}{2}(L + L^\top) \prec 0 $, then a trapping region exists.
To see this, rewrite the energy of the system \eqref{eq:Energy_LosslessSystem} in terms of the operator $L_S$ as
\begin{equation*}
    \frac{d}{dt} K_0\bigl(x(t)\bigr)
    \;=\; x(t)^\top c \;+\; x(t)^\top L_S\,x(t).
\end{equation*}
Then, it can be shown that if $L_s$ is \emph{negative definite}, there exists a finite radius outside which the energy decreases, thus implying system trajectories will eventually enter a finite ball from which they cannot exit---i.e.,~trajectories will remain bounded indefinitely. 
For a detailed derivation, we refer the reader to \cite{lorenz1963deterministic,Schlegel2015}.

Schlegel and Noack~\citep{Schlegel2015} further generalized this by  
leveraging the fact that quadratic systems preserve their nonlinearities under a coordinate shift, as described above in the introduction to Section \ref{sect_losslessquadROMS}. Schlegel and Noack showed that if there is a region such that the shifted operator $A_s(m)$ is negative definite, a less conservative estimate for a \emph{globally monotonically attracting} trapping region in the form of a ball $B\bigl(m,R_m\bigr)$ centered at $m$ with radius $R_m$ can be obtained. Moreover, Schlegel and Noack prove that the system \eqref{eq:sys_shifted} possesses a trapping region \emph{if and only if} there exists an $m$ such that
\begin{equation} \label{eq:LMI_boundedness}
    A_s(m) \;=\; L_s \;-\; \sum_{i=1}^n m_i\,Q^{(i)} \;\prec\; 0.
\end{equation}
In other words, by choosing $ m $ such that $ A_s(m) $ is negative definite, we ensure that the energy \eqref{eq:dKmdt} decreases outside the trapping region. When \eqref{eq:LMI_boundedness} holds, the radius of the trapping region centered at $m$ is given by
\begin{equation*}\label{eq:TR_SN_Radius}
    R_m \;=\; \frac{\|\;c + L\,m + f(m)\|\;}
         {\bigl|\lambda_{\max}\bigl(A_s(m)\bigr)\bigr|},
\end{equation*}
where $\lambda_{\max}\bigl(A_s(m)\bigr)$ is the largest (least negative) eigenvalue of $A_s(m)$.  The numerator $\|c+Lm+f(m)\|$ here essentially acts as a constant forcing or ``drift'' term, while the denominator $|\lambda_{\max}\bigl(A_s(m)\bigr)|$ is the least negative (i.e., worst‐case) rate at which the quadratic term can remove energy from the system. All trajectories of the system eventually enter $B(m,R_m)$ and remain there, establishing \emph{global boundedness}. As will be discussed in the next section, the radius $R_m$ is an upper-bound estimate on the trapping region size, and is therefore usually overly conservative.
For more detailed proofs and broader discussions of trapping regions in lossless quadratic systems, we refer the reader to~\cite{Schlegel2015,liao2024}.

\subsection{Optimal Trapping Regions via Convex Optimization}\label{sect:CVX_TrappingRegions}
While the Schlegel-Noack condition shown in equation  \eqref{eq:LMI_boundedness} gives a clear algebraic criterion for global boundedness, its practical implementation poses two challenges. First, verifying the trapping region condition requires a search over the translation vector $m\in\mathbb{R}^n$ to render the symmetric part $A_s(m)$ negative definite. \cite{Schlegel2015} cast this as a non-convex inf-sup problem, minimizing the maximal eigenvalue of $A_s(m)$, and advocate simulated annealing for its solution. However, without convexity there is no guarantee of finding the true global minimum; failure of the annealing search therefore leaves the boundedness test inconclusive, since a valid trapping region may still exist.

Second, when a candidate shift vector $m$ is found using the Schlegel and Noack method, the resulting radius shown in equation \eqref{eq:TR_SN_Radius} is an upper-bound approximation, and thus an inherently conservative overestimate. For a detailed explanation, we refer the reader to \cite{liao2024}. In short, this conservatism arises due to radius $R_m$ being estimated from the \emph{worst–case} alignment of the linear “drift” vector $c+Lm+f(m)$ with the maximum (most conservative) eigenvalue of the quadratic “dissipation” rate $\lambda_{max}(A_S(m))$. Consequently, the resulting bounds may be excessively large, which can be particularly problematic in reduced-order fluid dynamics models where unphysically large trapping regions do not accurately reflect the true system dynamics.

In our recent work~\citep{liao2024}, we developed a fully convex framework to remove these limitations and both certify the existence of trapping regions and compute their tightest possible radii for lossless quadratic systems. We provide a summary of these optimization problems here, and refer the reader to~\cite{liao2024} for detailed proofs and derivations.

To determine whether a trapping region exists for a given system, we recast the Schlegel-Noack criterion shown in equation \eqref{eq:LMI_boundedness} as a single linear matrix inequality (LMI). This is done by noting that in the condition \eqref{eq:LMI_boundedness}, the shifted term~$A_s(m)$ depends affinely on~$m$. Checking negative definiteness of $A_s(m)$ is thus an LMI feasibility problem~\citep{boyd2004convex}, which can be solved with standard convex optimization solvers (e.g., SeDuMi, MOSEK, SDPT3). Specifically, we introduce a scalar slack variable~$a$ and solve:
\begin{mini}
    {m \in \mathbb{R}^n, a \in \mathbb{R}}{a}{\label{SDP:min_lambda}}{}
    \addConstraint{A_s(m) \preceq a I_n.}
\end{mini}
If the optimal solution $a^* < 0$, then $A_s(m)\prec 0$ for some $m$, confirming that a trapping region exists. Conversely, if $a^* \geq 0$ then there are no coordinates under which the kinetic energy, $K_m(y)=\frac{1}{2}y^\top y$, verifies the existence of a trapping region for this system~\citep{liao2024}.

If a trapping region exists, we next refine the size of the trapping region. Instead of using the conservative estimate \eqref{eq:TR_SN_Radius}, we solve a quadratically constrained quadratic program (QCQP) to compute the exact minimal radius $R^*_m$:
\begin{subequations}  \label{eq:QCQP}
\begin{align}
    (R_m^*)^2 
    \;=\; 
    \max_{y} \;&\, y^\top y 
    \\[-6pt]
    \text{s.t.} \;&\, d(m)^\top y + y^\top A_s(m)\,y \;\ge\; 0.
\end{align}
\end{subequations}
Although this QCQP is non-convex, the dual function of its Lagrangian form admits a convex semi-definite programming~(SDP) dual problem~\eqref{eq:QCQPdual} that can be solved to global optimality \citep{boyd2004convex}
\begin{subequations} \label{eq:QCQPdual}
\begin{align}
    (R_m^*)^2 \;=\;
    \min_{\lambda \geq 0,\beta} 
    & \;\beta 
    \\[-6pt]
    \text{s.t.}
    & \;\begin{bmatrix}
        I_n \;+\; \lambda\,A_s(m) & \;\tfrac{\lambda}{2}\,d(m) \\
        \tfrac{\lambda}{2}\,d(m)^\top & -\beta
    \end{bmatrix}
    \;\preceq\; 0.
\end{align}
\end{subequations}
Here, $\lambda$ is the primal-Lagrangian variable, and $\beta$ is a dual variable. Solving \eqref{eq:QCQPdual} yields $R_m^*$, the radius of the tightest possible trapping region $B(m,R_m^*)$. By leveraging this convex formulation, we avoid the conservatism of the original approach, enabling more precise and physically realistic bounds on the system’s dynamics. 

Using several numerical examples, we demonstrated in \citep{liao2024} that both the trapping region existence certificate and the  optimal radius can be solved efficiently with standard SDP solvers, yielding bounds often orders of magnitude tighter than prior analytical estimates. These convex formulations form the backbone of the present data‐driven identification loop, ensuring that for any choice of hyperparameters we can both \emph{guarantee} whether a lossless, globally bounded ROM exists and, if it does, compute its \emph{optimal} trapping region in a single unified framework. 

\subsection{Identification of Sparse and Lossless ROMs from Data}\label{sec:SINDy}

The ability to certifiably determine optimal trapping regions---as discussed in Section~\ref{sect:CVX_TrappingRegions}---can be leveraged within data-driven modeling frameworks, such as the operator inference framework~\citep{Peherstorfer2016, goyal2024,kramerARFM2024} and the sparse identification of nonlinear dynamics~(SINDy) framework~\citep{Brunton_SINDy,Loiseau2018}.
In this work, we demonstrate integration within the SINDy modeling framework, which is an optimization-based regression algorithm commonly used for identifying sparse governing equations. 
The sparsity-promoting nature of SINDy enables the construction of more robust models despite errors in finite datasets, as promoting sparsity tends to prevent error propagation through coupled state variables.

In SINDy, we first collect $n_t$ ``snapshots'' of the time evolution of the $n$-dimensional state $x$ and its time-derivative $\dot{x}$ into data matrices $ {X} \in \mathbb{R}^{n_t \times n} $ and $ {\dot{X}} \in \mathbb{R}^{n_t \times n}$, respectively.
Each column of $X$ corresponds to the time series of an individual state variable $ x_i(t) $ sampled at discrete time points $ t_1, t_2, \dots, t_{n_t} $, and similarly for $\dot{X}$.
In this work, we determine $\dot{X}$ from the data in $X$ using an eighth-order central difference scheme.
The data matrices $X$ and $\dot{X}$ are then related by the dynamics, and SINDy provides a convenient way to express the right-hand side of \eqref{eq:Dynamic_System} as the product of a \emph{candidate function library $\Theta(X)$} and a \emph{coefficient matrix $\Xi$}:
\begin{equation} \label{eq:SINDy}
    \dot{X} = {\Theta}(X) \, {\Xi}, 
    \quad \quad
    \Xi\in \mathbb{R}^{  (1+n+\frac{n(n+1)}{2})\times n}, 
    \quad 
    \Theta(X) \in \mathbb{R}^{n_t \times (1+n+\frac{n(n+1)}{2})}.
\quad
\quad
\end{equation}
The candidate function library ${\Theta}(X)$ is a data matrix consisting of constant (affine), linear, and quadratic functions of the columns of $X$ as needed to describe the flow.
The candidate function library is constructed as
\begin{equation} \label{Theta_X}
\Theta(X)=
    \begin{bmatrix} 
    \vdots & \vdots & \vdots \\
    1 & X^{P_1} & X^{P_2}  \\
    \vdots & \vdots & \vdots \\
    \end{bmatrix},
\end{equation}
where the first column represents the affine terms, $ X^{P_1} $ contains linear functions of the data in $X$, and $ X^{P_2} $ contains quadratic functions of the data in $X$.
Note that we have chosen $ X^{P_2}$ to contain the upper-triangular (non-redundant) entries of the symmetric quadratic tensors $ Q^{(i)} $ (recall the definition of $f$ in \eqref{eq:Dynamic_System}), arranged such that each row corresponds to a unique monomial $ x_j x_k $ with $ j \leq k $. To correctly represent the symmetric quadratic form $ {x}^\top Q^{(i)} {x} $, the off-diagonal terms $ x_j x_k $ for $ j < k $ in $ X^{P_2}$ are scaled by a factor of 2, since each cross term appears twice in the full expansion. 
The associated coefficient matrix $\Xi$ is defined as,
\begin{equation}
{\Xi} 
\;=\;
\begin{pmatrix}
c^\top\\
L \\
Q^{\triangle} \\
\end{pmatrix},
\quad
c \,\in\, \mathbb{R}^{n}, 
\quad
\quad
L \,\in\, \mathbb{R}^{n \times n}, 
\quad
Q^{\triangle} \,\in\, \mathbb{R}^{(\frac{n(n+1)}{2})\times n},
\end{equation}
and includes the coefficients associated with the affine terms $c$ , linear terms $L$, and quadratic terms $ Q^{\triangle}$.

The coefficient matrix $\Xi$ in \eqref{eq:SINDy} is unknown, but can be determined via regression.
The SINDy framework exploits the fact that only a few terms in $\Theta(X)$ will contribute to the dynamics of the flow, and so seeks to find a \emph{sparse} coefficient matrix $\Xi$ that best represents the dynamics.
We note that even if the true underlying equations of motion are not sparse, promoting sparsity has the advantage of introducing model robustness to truncation errors and other factors.
In addition to promoting sparsity in $\Xi$, we must ensure that the resulting model will possess quadratic terms that are \emph{lossless}, which can be achieved by solving for a coefficient matrix ${\Xi}$ that  enforces $x^\top f(x)=0$.
To this end, we define a constrained least-squares regression  problem with an $L_1$-regularization (sparsity-promoting) term as~\citep{Loiseau2018},
\begin{mini}
  {{\Xi}}{\left\| {\Theta}(X) \,{\Xi} - \dot{X} \right\|_2^2+\delta \|\Xi\|_1}{}{}
  \addConstraint{{C} \,{\Xi}(:) \;=\; {d}}.
  \label{eq:SINDyConstrained}
\end{mini}
Here, the term $\delta \|\Xi\|_1$ promotes sparsity in $\Xi$, with the parameter $ \delta $ controlling the degree to which sparsity is promoted. 
A larger $ \delta $ promotes a sparser solution, while a smaller $ \delta $ results in more active terms in $ \Xi$; in general, the degree of sparsity with respect to $\delta$ is problem dependent and cannot be determined \emph{a priori}. The linear constraint $C\Xi(:)=d$, where $\Xi(:)=\xi$ is the vectorized form of the matrix $\Xi$, ensures the identified system is lossless by encoding algebraic constraints on the quadratic entries of $\Xi(:)$ via the matrix ${C}$ and vector ${d}$. For an explanation of how the constraint matrix $C$ and the vector $d$ are determined, refer to the Appendix \ref{apdx:triadic}. 

By jointly enforcing sparsity and the lossless property, SINDy can be used to find a model that is physically consistent with governing incompressible Navier-Stokes equations. However, enforcing the nonlinear terms to be lossless is not sufficient to guarantee that resulting model trajectories will be long-time bounded.
Thus, the resulting model may not be \emph{dynamically consistent} with respect to the flow being modeled.
For incompressible flows that admit a trapping region, matching the optimal trapping region size provides a means of obtaining models that are (at least approximately) dynamically consistent.
In the next section, we describe how the convex optimization problems associated with optimal trapping regions from Section~\ref{sect:CVX_TrappingRegions} can be integrated within the SINDy framework. 

\section{Optimally Bounded Nonlinear Models from Data}\label{sec:methodMain}
\label{sect:Algorithm}

We now describe how to combine the convex optimization methods described in \ref{sect:CVX_TrappingRegions} with the data-driven identification of a sparse, lossless, and \emph{bounded} quadratic reduced-order model via SINDy, as described in section \ref{sec:SINDy}.
To begin, we will represent the system~\eqref{eq:Dynamic_System} in terms of a candidate function library and coefficient matrix, as in the SINDy formulation~\eqref{eq:SINDy}. We then modify the constrained least-squares problem~\eqref{eq:SINDyConstrained} to simultaneously enforce boundedness, sparsity, and losslessness in the identified dynamics. 

Our objective is to find the coefficient matrix ${\Xi}$ such that the resulting model is sparse, conserves energy via lossless quadratic terms, and admits a provably optimal trapping region that guarantees long-time boundedness of model trajectories. 
In addition to solving for the coefficient matrix $ \Xi $, we must also find the shift vector $ m $. Recall from section \ref{sect:SystemBoundedness}, that by shifting the system to be centered at $ m $ such that $ A_S(m) $ is negative definite, we ensure that the energy \eqref{eq:dKmdt} decreases outside the trapping region. 
As discussed in section \ref{sect:CVX_TrappingRegions}, we certify this condition in a single convex program by introducing a scalar slack variable $a$ to relax the strict negative‐definiteness condition into a tractable inequality.
Thus, for a given coordinate shift $m$, we seek the trapping region $B(m,R_m)$ centered about $m$ that has the smallest radius. 
This is formulated as an optimization problem that seeks to minimize the difference between the model prediction and observed data, while promoting a sparse solution for $ \Xi $ and enforcing system stability. The optimization problem is defined as
\begin{mini} 
    {\Xi,m, a}{\|\Theta(X) \Xi - \dot{X}\|_2^2 + \delta \|\Xi\|_1 + a + \eta\,\|m\|_2}{}{}
    \addConstraint{C\Xi(:)=d}
    \addConstraint{A_S(m, \Xi) \preceq  a I_n}
    \addConstraint{-\gamma \leq  a.}\label{eq:TrOPTIMALEqn}
\end{mini}
Just as in the SINDy equation \eqref{eq:SINDyConstrained}, first term of the objective function is the least-squares error between the predicted dynamics $ \Theta(X) \Xi $ and the observed time derivatives $ \dot{X} $, while the second term is an $ \ell_1 $-regularization term that promotes sparsity in $ \Xi $. The third term, $a$ is the previously discussed slack variable for the trapping region condition. The final term in the cost function $\eta \|m\|_2$ is added to regularize the problem and ensure uniqueness when multiple $m$ satisfy the trapping condition.
The constraint $C\Xi(:) = d$, ensures that the nonlinearity is lossless. The second constraint $A_S(m, \Xi) \preceq a I_n$ along with the minimizing $a$ ensures boundedness by requiring the matrix $ A_S(m, \Xi) $ to be optimally negative definite.  
Note that the term $A_S(m,\Xi)$ has been modified from \eqref{eq:As} to be a function of \emph{both} $m$ and $\Xi$. This notation is introduced because $A_S(m)$ is dependent on the linear term $L$ and quadratic terms $Q^{(i)}$, which are now encoded in the decision variable $\Xi$. 
Finally, to prevent the optimization objective function from being unbounded from below, we introduce the constraint $-\gamma \leq a$.
The positive scalar parameter $ \gamma $ controls how much stability is enforced by defining the dissipation rate; larger values of $\gamma$ allow the semidefinite constraint on the dissipation term $A_S(m,\Xi)$ to be  \emph{more} negative definite, implying a stronger boundedness property.

\subsection{The Trapping-SDP Modeling Framework}
The Trapping-SDP procedure presented here solves for both the coefficient matrix $ \Xi $ and the shift vector $ m $ while promoting sparsity, boundedness, and enforcing losslessness in the nonlinear term. To solve the problem using convex optimization techniques, we consider two alternating steps: a regression step for the coefficient matrix $ \Xi $ and a stability-enforcement step for the shift-vector $ m $ and a slack variable $ a $.
This is because the full optimization problem involves minimizing the least-squares error, promoting sparsity through the $ \ell_1 $-norm, constraining the nonlinearity to ensure losslessness,
and enforcing boundedness via a semidefinite constraint. Solving all these requirements in one step would create a complex optimization problem with competing objectives, making it difficult to handle efficiently. By decomposing the problem into two subproblems, we can use readily available convex optimization solvers for each step.
This alternating minimization approach---sometimes referred to as \emph{block coordinate descent}---is commonly used for problems where separate components (in this case, sparsity and stability) influence different parts of the optimization problem. It simplifies each subproblem and accelerates convergence. The procedure proceeds in the following steps:
\begin{enumerate}
    \item {Variable initialization}
    \item {Regression}
    \item {Stability-enforcement}
    \item {Iteration of steps} (i) {and} (ii)
    \item {Sparsity refinement}.
\end{enumerate}
This framework depends on two hyperparameters: the sparsity-promoting weight $\delta$ and the stability-enforcement term $\gamma$.
We are interested in determining a model that best balances the desire for small modeling error, a large degree of sparsity, and a small trapping region.
Therefore, we sweep over a range of values for $\delta$ and $\gamma$, repeating the modeling procedure described above for each parameter pair, and then selecting the model that best balances the desired features.

For each candidate model resulting from the parameter sweep, we evaluate the performance using a validation-based procedure. Specifically, we first train the model on a finite trajectory from the training data. Then, we generate validation trajectories by time-marching the identified model from a set of off-training initial conditions, ensuring the model is validated against data that was not seen during the training phase.

The model's accuracy is then assessed by computing the root mean square (RMS) error between the predicted dynamics $x_i(t)$ and the validation data $x_{i,true}(t)$
\begin{equation}
    RMSE=\sqrt{\frac{1}{n}\sum_{i=1}^{n}\bigl(x_i(t)-x_{i,true}(t)\bigr)^2}.
\end{equation}
We compute the trapping region associated with each model and record its radius.
As a measure of model sparsity, we use the sparsity ratio: i.e.,~the ratio of the number zero elements to the total number of elements in $\Xi$.
The final model to be selected is the one that best balances small prediction error, large sparsity ratio, and small trapping region radius. In practice, we observe that models with the lowest prediction error often also possess the tightest or nearly-tightest trapping regions. This multi-criterion model-selection approach ensures that the final model is accurate, robust to numerical error, and globally bounded.

We will now discuss each step of the procedure in greater detail.

\subsubsection{Initialization}\label{subsec:M_Init}
It is necessary to first test whether a globally attracting trapping region can exist for the provided training data and hyperparameter $\gamma$. 
We therefore begin by computing an unconstrained least‐squares estimate of the coefficient library $\Xi^{(0)}$, which provides an approximation of the dynamics in the chosen function library $\Theta(X)$:
\begin{equation}
\begin{aligned}
\Xi^{(0)} \;\gets\; &\arg\min_{\Xi}\;\tfrac{1}{2}\,\|\Theta(X)\,\Xi \;-\;\dot{X}\|_2^2,\\
&\text{subject to}\quad C\,\Xi(:) = d.
\end{aligned}
\end{equation}
This initial fit ensures that the initial coefficient matrix $\Xi^{(0)}$ lies in a region of the parameter space that accurately reflects the underlying physics, rather than an arbitrary model. Furthermore, solving a least‐squares problem with the lossless constraints $C\,\xi = d$ preserves any known symmetry or conservation laws from the outset, preventing later iterations from chasing infeasible or physically inconsistent solutions. Finally, this initialization provides a well‐conditioned starting point for the semidefinite programming step: since $\Xi^{(0)}$ already captures the principal dynamics, the subsequent searches for a trapping region via coordinate shift $m$ and slack variable $a$ are more likely to find a valid bounded solution, rather than failing due to an ill‐posed initial guess. In practice, this regression‐first strategy greatly accelerates convergence of the alternating optimization and enhances the robustness of the final sparse, lossless, and globally bounded ROM. 

Once the coefficient matrix $\Xi^{(0)}$ is obtained, it is necessary to verify whether the current model admits a globally attracting trapping region and, if so, to compute an initial estimate of its center.  By the Schlegel–Noack trapping theorem, there exists a trapping region if and only if one can find a shift vector $m\in\mathbb{R}^n$ such that the shifted symmetric operator \[A_S(m) \;=\; L_S(\Xi^{(0)}) \;-\;\sum_{i=1}^n m_i\,Q^{(i)}(\Xi^{(0)})\] is negative definite.  
As discussed in section \ref{sect:CVX_TrappingRegions}, we certify this condition in a single convex program by using the slack variable $a$ to relax the strict negative‐definiteness condition into a tractable inequality
\begin{equation}
\begin{aligned}
(m^{(0)},\,a^{(0)}) \;\gets\;&\arg\min_{m,\,a}\;\bigl(a + \eta\,\|m\|_2\bigr),\\
&\text{subject to}\quad A_S\bigl(m,\Xi^{(0)}\bigr)\;\preceq\; a\,I_n,\\
&\qquad\quad -\gamma \;\le\; a.
\end{aligned}
\end{equation}
As in equation \eqref{eq:TrOPTIMALEqn}, the constraint $-\gamma \leq a$, prevents the SDP from being unbounded from below. The regularization term in the cost function $\eta \|m\|_2$ helps condition this SDP, as there could exist multiple shift vectors $m$ that satisfy the first constraint. The regularization term $\eta\|m\|_2$ biases the solution towards smaller shifts, improving conditioning and ensuring uniqueness when multiple $m$ satisfy the trapping condition.
Upon convergence, a negative optimal value $a^{(0)}<0$ provides a rigorous certificate that a globally attracting trapping region exists for the initial ROM, allowing the initial estimates $(m^{(0)},a^{(0)})$ to then seed the alternating regression and stability-enforcement iterations that jointly refine both $\Xi$ and the trapping region parameters. Conversely, in the case that $a^{(0)}\ge0$, we can definitively conclude that a globally attracting trapping region does not exist for the current model coefficients, and so can discard the model. 

This two‐step initialization is both theoretically grounded and practically effective: the least‐squares fit captures the dominant dynamics from data, and the SDP then determines whether global boundedness is attainable.  Together, they provide a  starting point for seeking a model with an \emph{optimal} trapping region.

\subsubsection{Iterative Regression and Stability-Enforcement}\label{subsec:MRegStab}
We next proceed with an alternating block-coordinate descent loop, updating the coefficients and the stability certificate in turn until both have converged. For the regression step, we solve the following convex optimization problem to update $ \Xi $ at each iteration $ k $, with $ m^{(k-1)} $ and $ a^{(k-1)} $ fixed:
\begin{equation}\label{eq:XiStep}
\begin{aligned}
\Xi^{(k)} \gets &\arg\min_{\Xi} && \|\Theta(X) \Xi - \dot{X}\|_2^2 + \delta \|\Xi\|_1 \\
& \: \text{s.t.} && C\Xi(:) = d \\
& && A_S(m^{(k-1)}, \Xi) \preceq a^{(k-1)}I_n. 
\end{aligned}
\end{equation}
Here, the $\ell_1$-penalty $\delta\|\Xi\|_1$ promotes sparsity in the coefficient matrix $ \Xi $, while the semidefinite constraint enforces that the current linear and quadratic operators admit a trapping region when shifted by $m^{(k-1)}$.  Because this is a convex program, we can obtain the updated estimate for $\Xi^{(k)}$ efficiently. Next, holding $\Xi^{(k)}$ fixed, we refine the shift vector $m$ and slack variable $a$ by solving
\begin{equation} \label{eq:SDP_m_a}
\begin{aligned}
m^{(k)}, a^{(k)} \gets &\arg\min_{m, a} && a + \eta \|m\|_2 \\
& \: \text{s.t.} && A_S(m, \Xi^{(k)}) \preceq a I_n \\
 & && -\gamma \leq a.
\end{aligned}
\end{equation}
This SDP updates the trapping-region center $m$ and slack variable $a$ so that the shifted symmetric operator $A_S(m, \Xi^{(k)})$ remains negative definite, thereby enforcing global boundedness for the current model. Again, we use a small $\eta$ so that the $\eta\|m\|_2$ term regularizes the shift so as to avoid ill-conditioned solutions. We iterate between the regression and stability-enforcement steps until convergence is achieved, based on the following criteria:
\begin{equation*}
\bigl\lvert\Xi^{(k)} - \Xi^{(k-1)}\bigr\rvert \;\le\; \epsilon_{\Xi}
\quad\text{and}\quad
\bigl\lvert m^{(k)} - m^{(k-1)}\bigr\rvert \;\le\; \epsilon_{m}.
\end{equation*}
At convergence, the trapping-region condition is checked. If $a^{(k)}\geq 0$, the identified model is not guaranteed to be bounded and thus removed as a candidate.
If $a^{(k)}<0$, we obtain a model that simultaneously fits the data, is sparse as prescribed by the hyperparameter~$\delta$, and provably admits an optimal trapping region.  This alternating scheme leverages the convexity of each subproblem to ensure efficient and reliable computation of the final ROM.
\subsubsection{Convergence and Refinement}\label{subsec:MRefine}

We will now briefly discuss the convergence properties of the algorithm and the final refinement step. Let
\begin{equation}
    J(\Xi^{(k)},m^{(k)},a^{(k)}) \;=\; \tfrac12\|\Theta(X)\,\Xi^{(k)} - \dot X\|_2^2 \;+\;\delta\|\Xi^{(k)}\|_1  + a^{(k)} + \eta \|m^{(k)}\|_2 
\end{equation}
denote the objective function at iteration $k$. 
We note that the objective is bounded from below by $0$. At the initial iterate ($k=0$), the objective function is $J(\Xi^{(0)},m^{(0)},a^{(0)})$
by virtue of initialization. 
At each subsequent iterate $k$,
\begin{equation}
J(\Xi^{(k-1)},m^{(k-1)},a^{(k-1)})\geq J(\Xi^{(k)},m^{(k-1)},a^{(k-1)}) \geq J(\Xi^{(k)},m^{(k)},a^{(k)}),
\end{equation}
by virtue of convexity. Because each substep cannot increase the objective and $J(\Xi,m,a)\geq -\gamma$ is bounded below with the constraint $-\gamma \leq  a$, the sequence of costs is non-increasing and converges to a finite value.

After iterating through the regression and stability steps, we check for convergence based on the changes in $ \Xi $ and $ m $. The algorithm terminates when both updates fall below the prescribed tolerances $ \epsilon_{\Xi} $ and $ \epsilon_m $.
Once convergence is achieved, we refine the solution by fixing the sparsity pattern and forcing small terms in $ \Xi $ to zero. Let $ \Xi^{K} $ be the converged coefficient vector, and define a tolerance such that any $ |\Xi^{K}_i| < \epsilon_x $ is considered small. We construct the matrices $ C_{\text{ref}} $ and $ d_{\text{ref}} $ as:
\[
C_{\text{ref}}(i,j) = 
\begin{cases}
1, & \text{if } |\Xi^{K}_j| < \epsilon_x \\
0, & \text{otherwise}
\end{cases}
\quad \text{and} \quad d_{\text{ref}} = \mathbf{0}.
\]
The augmented constraint matrices are
\[
\tilde{C} = \begin{pmatrix} C \\ C_{\text{ref}} \end{pmatrix}, \quad \tilde{d} = \begin{pmatrix} d \\ d_{\text{ref}} \end{pmatrix}.
\]
Using the optimal solutions for the shift vector $m^{*}$ and the slack variable $a^{*}$, obtained in the last iteration of \eqref{eq:SDP_m_a}, we then solve the optimization problem one final time with these updated constraints to ensure the small terms are forced to zero:
\begin{equation}
\begin{aligned}
\Xi^{*} \gets &\arg\min_{\Xi} && \|\Theta(X) \Xi - \dot{X}\|_2^2 + \delta \|\Xi\|_1 \\
& \: \text{s.t.} && \tilde{C}\Xi = \tilde{d} \\
& && A_S(m^{*}, \Xi) \preceq a^{*} I_n. 
\end{aligned}
\end{equation}
This step finalizes the sparse and bounded model by eliminating negligible coefficients. We note that the refinement step could increase the cost
\begin{equation}
    J(\Xi^{(k-1)},m^{(k-1)},a^{(k-1)})\leq J^{*}(\Xi^{*},m^*,a^*)
\end{equation}
as more constraints are imposed during this step. However, since we terminate when both
$\|\Xi^{(k)}-\Xi^{(k-1)}\|$ and $\|(m^{(k)},a^{(k)})-(m^{(k-1)},a^{(k-1)})\|$ fall below user‐specified tolerances, we ensure a self‐consistent, locally optimal solution.

\section{Numerical Benchmark Examples}\label{sec:numericalEx}
In this section, we provide two numerical benchmark examples to illustrate the proposed method on systems with known dynamics: (1)~a Lorenz dynamical system~\citep{lorenz1963deterministic}, and (2)~a reduced-order model of a sinusoidally forced shear flow~\citep{Moehlis2004}.
Application to an unsteady nonlinear aerodynamics problem follows in section~\ref{sec:AirfoilResults}.

\subsection{Lorenz System}

To illustrate our approach, we first examine the following Lorenz dynamical system~\citep{lorenz1963deterministic}:
\begin{subequations} \label{eq:Lorenz}
    \begin{align}
        \frac{d x_1}{dt} &= -\sigma x_1 + \sigma x_2, \\
        \frac{d x_2}{dt} &= \rho x_1 - x_2 - x_1 x_3, \\
        \frac{d x_3}{dt} &= -\alpha x_3 + x_1 x_2,
    \end{align}
\end{subequations}
with parameters $\sigma = 10$, $\rho = 28$, and $\alpha = \frac{8}{3}\approx2.667$. Although the Lorenz system exhibits chaotic behavior and lacks stable equilibrium points, it has been shown to be bounded~\citep{Schlegel2015}. In our analysis, we verify that by applying a translation~$m$, the symmetric part of the linear operator, $A_s(m)$, becomes negative definite. According to the theorem discussed in section \ref{sect:SystemBoundedness}, this implies the existence of a trapping region $B(m, R_m)$ with an estimated radius $R_m = 101.33$. We note that this radius slightly differs from that reported in \cite{Schlegel2015}, due to a correction noted by~\cite{Kaptanoglu2021}.

In our previous work~\citep{liao2024}, we used our Trapping-SDP method to determine a trapping region for the Lorenz system. There, we obtained a translation $m = [0,\, 0,\, 38]^\top$ and an optimal trapping region radius of $R_m^* = 39.25$, which is significantly smaller than the conservative bound of $R_m = 101.33$. It is important to note that these results were achieved with full knowledge of the system dynamics. In contrast, our current objective is to employ our data-driven algorithm to recover the dynamic equations from a sample trajectory of the Lorenz system without prior knowledge of the governing equations, and to obtain a trapping region size close to the optimal value determined previously.

\begin{figure}
\centering
\includegraphics[width=0.8\textwidth]{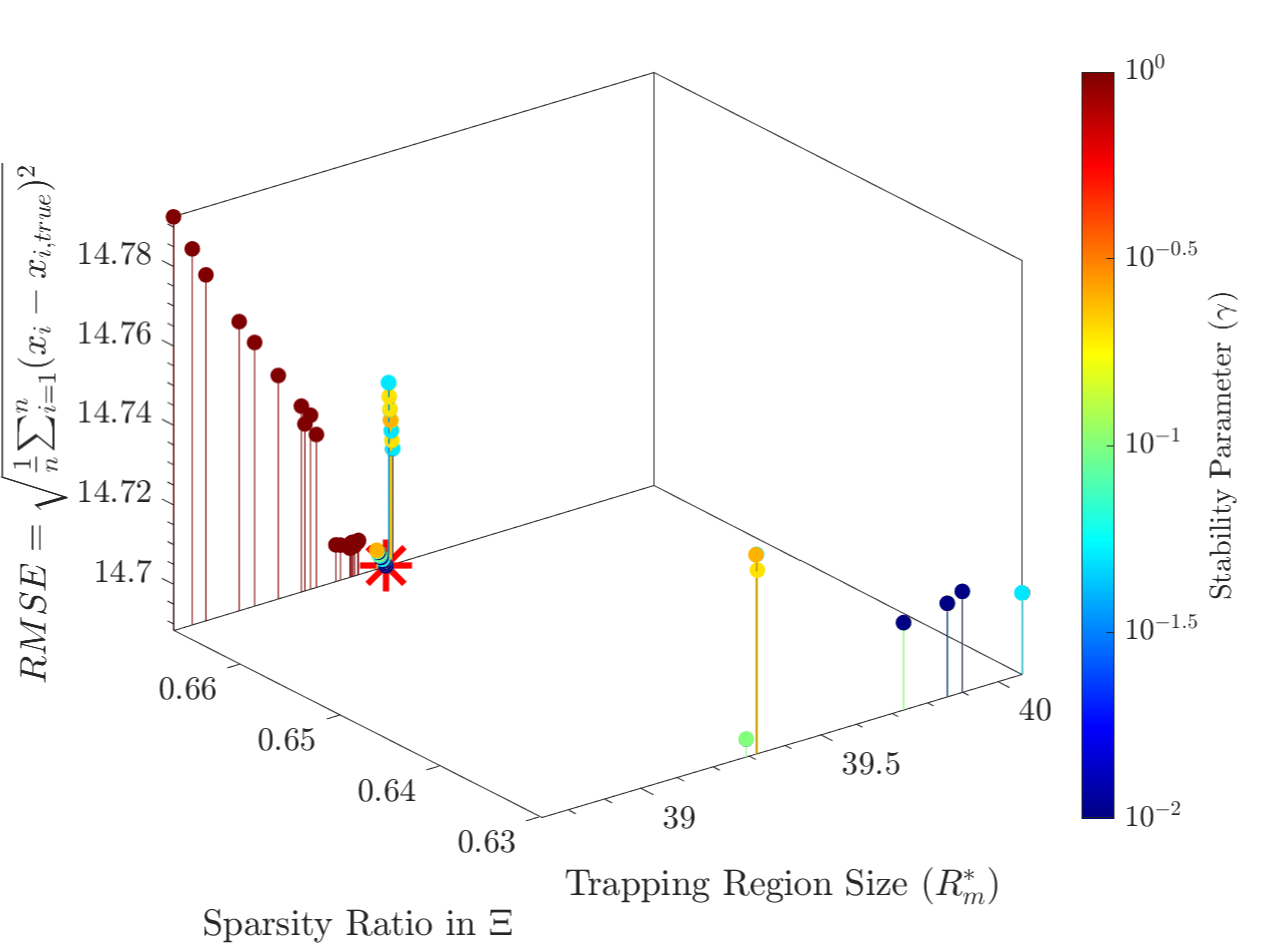}
\caption{Model selection for the three‐state Lorenz ROM. Each marker denotes a candidate model obtained by sweeping the sparsity parameter $\delta$ (horizontal axis: sparsity ratio in $\Xi$) across two levels and varying the stability weight $\gamma$ (color scale). The trapping region size $R_m^*$ (depth axis) and the root‐mean‐square error (RMSE) of the state predictions (vertical axis) are shown for each model. Because the Lorenz system has only three states, only two sparsity patterns arise, producing two vertical “stems” of points. The denser stem (lower sparsity) yields larger RMSE and looser bounds, while the sparser stem (higher sparsity) achieves both lower error and a tighter trapping region. The optimal model, indicated by the red star, corresponds to the higher‐sparsity solution with the smallest RMSE and trapping region.}
\label{fig:Error3d_Lorenz}
\end{figure}

To this end, we integrated the Lorenz equations~\eqref{eq:Lorenz} forward in time over 5000 time steps from a random initial condition. We then applied the algorithm described in section~\ref{sect:Algorithm}, sweeping over a range of sparsity ($\lambda$) and stability ($\gamma$) parameters, ultimately choosing a model that balanced accuracy and trapping region size (as shown in figure \ref{fig:Error3d_Lorenz}). Our method successfully recovered the sparsity pattern of the original Lorenz system with parameters $\sigma = 9.9903$, $\rho = 27.9684$, and $\alpha = 2.6661$. The trapping region for the identified model was $R_m^* = 39.3144$, closely matching the optimal value from our earlier work~\citep{liao2024}. These results are summarized in table \ref{tab:LorenzResults}.

\begin{table*}
\centering
\caption{\label{tab:LorenzResults}Tabulated comparison of the true Lorenz parameters used to generate the training trajectory with the parameters obtained using our method (labeled as ``model result''). The table also compares the optimal trapping region of the true Lorenz system with the trapping region of the model.}
\begin{tabular}{l@{\hskip 0.25in}c@{\hskip 0.25in} c@{\hskip 0.25in} c@{\hskip 0.25in} c@{\hskip 0.25in}}
\textbf{Parameter}       & \textbf{$\sigma$}   & \textbf{$\rho$}  & \textbf{$\alpha$}  & \textbf{Trapping Region size} $R_m^*$ \\ \hline
    \\ \textbf{Lorenz} & $10$  & $28$   & $\frac{8}{3}\approx2.667$ & $39.25$ \\ \hline
    \\ \textbf{Model Result} & $9.9903$ & $27.9684$ & $2.6661$ & $39.3144$\\ \hline
\end{tabular}
\end{table*}

\begin{figure}
\centering
\includegraphics[width=0.9\textwidth]{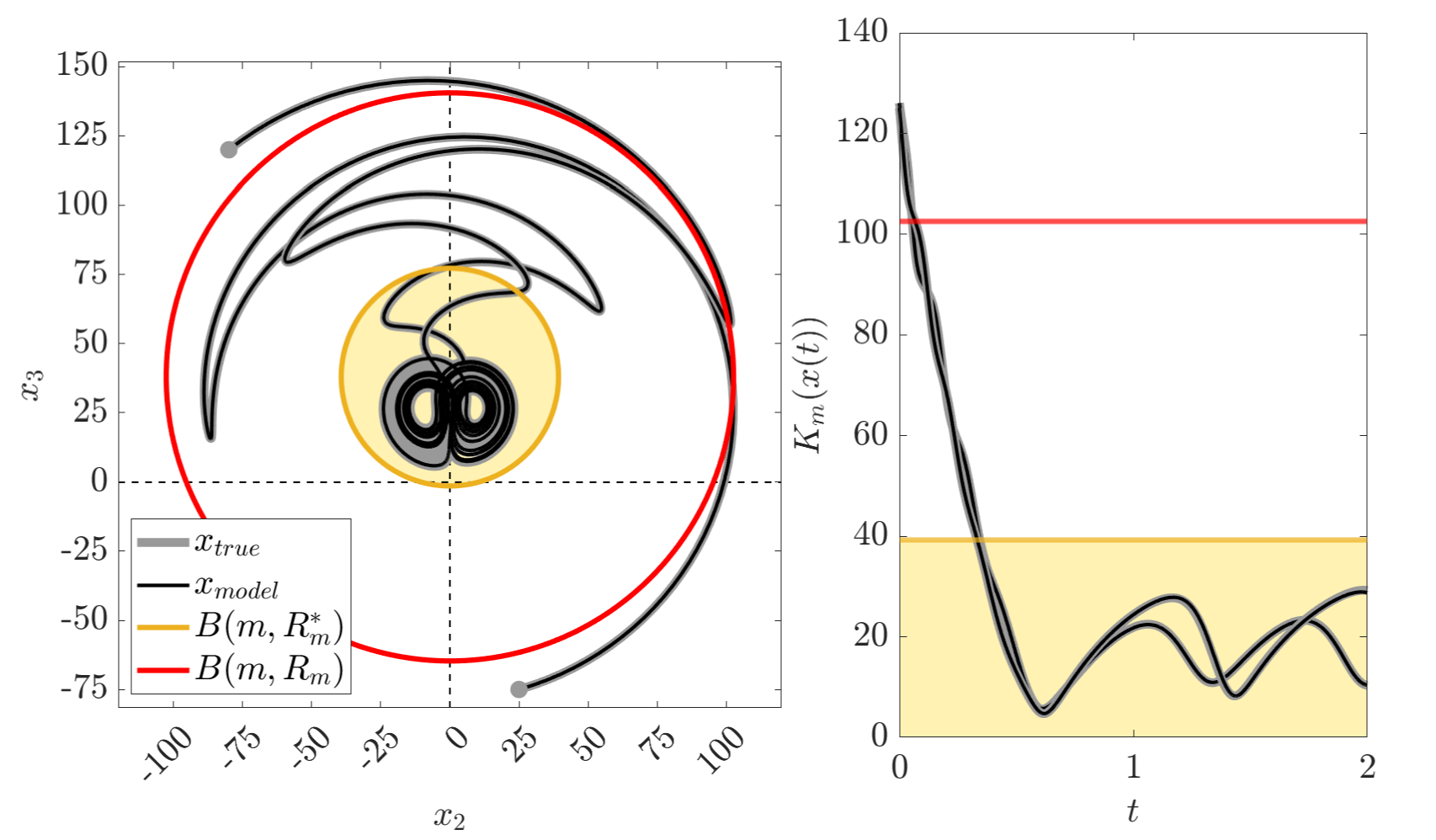}
\caption{Trapping region comparison. $B(m,R^*_m)$ shows the trapping region computed via our method, and $B(m,R_m)$ shows the trapping region computed via the Schlegel and Noack method. }
\label{fig:Lorenze}
\end{figure}

To validate the results, we integrated the discovered model from a set of initial conditions not used for training and compared them to the true dynamics. These results are summarized in figure \ref{fig:Lorenze}, where we also compare the trapping region identified using our method with that of Schlegel and Noack. Note that the energy of the shifted system $K_m(x(t))$ decays monotonically towards the trapping region. 

\subsection{Sinusoidally-Forced Shear Flow}

Next, we consider the 9-state Galerkin model of a sinusoidally forced shear flow introduced by~\cite{Moehlis2004}, which has been used in a number of previous works to demonstrate nonlinear stability analysis methods (see, e.g.,~\cite{Goulart2012,Liu2020,Kalur2022,PhysRevFluids.10.014401,enayati2025,buzhard2025}).  
The flow consists of a sinusoidal body force applied between two free‑slip, counter‑moving walls. 
The reduced-order Galerkin model is based on a projection of the dynamics onto nine Fourier modes, each capturing a physically important feature of the flow dynamics.
For more details on the model and its physical significance, the reader is referred to the original source~\citep{Moehlis2004}.

The 9-state model has the form
\begin{equation} \label{eq:9state}
\dot{\tilde{x}} = f(\tilde{x};Re) = \tilde{L}(Re)\tilde{x}+ Q(\tilde{x})\tilde{x},
\end{equation}
where the state vector $\tilde{x}=\tilde{x}(t)\in\mathbb{R}^{9}$ consists of nine modal coefficients which govern the temporal evolution of the flow. The quadratic term is defined as
\begin{equation}
    Q(\tilde{x})\,\tilde{x} \;=\;
    \begin{bmatrix}
        \tilde{x}^\top Q^{(1)} \tilde{x} \\[3pt]
        \vdots \\[3pt]
        \tilde{x}^\top Q^{(9)} \tilde{x}
    \end{bmatrix},
\end{equation}
where each $Q^{(i)}\in\mathbb{R}^{9\times9}$ is symmetric and  $\tilde{x}^\top Q(\tilde{x})\,\tilde{x}\equiv0$, i.e. the quadratic nonlinearity is lossless.  The linear operator $\tilde{L}(Re)$ is Hurwitz, and is parameterized by the Reynolds number ${Re}>0$. In the original formulation the nontrivial equilibrium sits at 
\[
  \tilde x = [\,1,\,0,\dots,0\,]^\top,\quad
  \dot{\tilde x}=0.
\]
To recenter this fixed point at the origin, we introduce the shift 
\[
  x = \tilde x - c,\qquad
  c = [\,1,\,0,\dots,0\,]^\top,
\]
so that $\dot x = 0$ at $x=0$.  This translation modifies the linear operator: if $\widetilde L(Re)$ is the original linear term, then
\[
  L(Re) \;=\; \widetilde L(Re)\;+\;W,
\]
where $W\in\mathbb{R}^{9\times9}$ is chosen to absorb the cross terms introduced by the shift, namely
\[
  W\,\tilde x \;=\; Q(\tilde x)\,c \;+\; Q(c)\,\tilde x,
\]
with $Q(\cdot)$ being the collection of symmetric quadratic operators.  

We apply our modeling framework for a Reynolds number of $Re=200$ to provide a stringent test of our method’s ability to recover both the correct governing coefficients and a certifiably tight trapping region.
As noted by \cite{Moehlis2004}, the case of $Re=200$ results in strongly nonlinear interactions with intermittent bursts.
The strong energy exchanges and modal interactions in this flow make obtaining accurate and bounded ROMs from data challenging.

\begin{figure}
\centering
\includegraphics[width=0.8\textwidth]{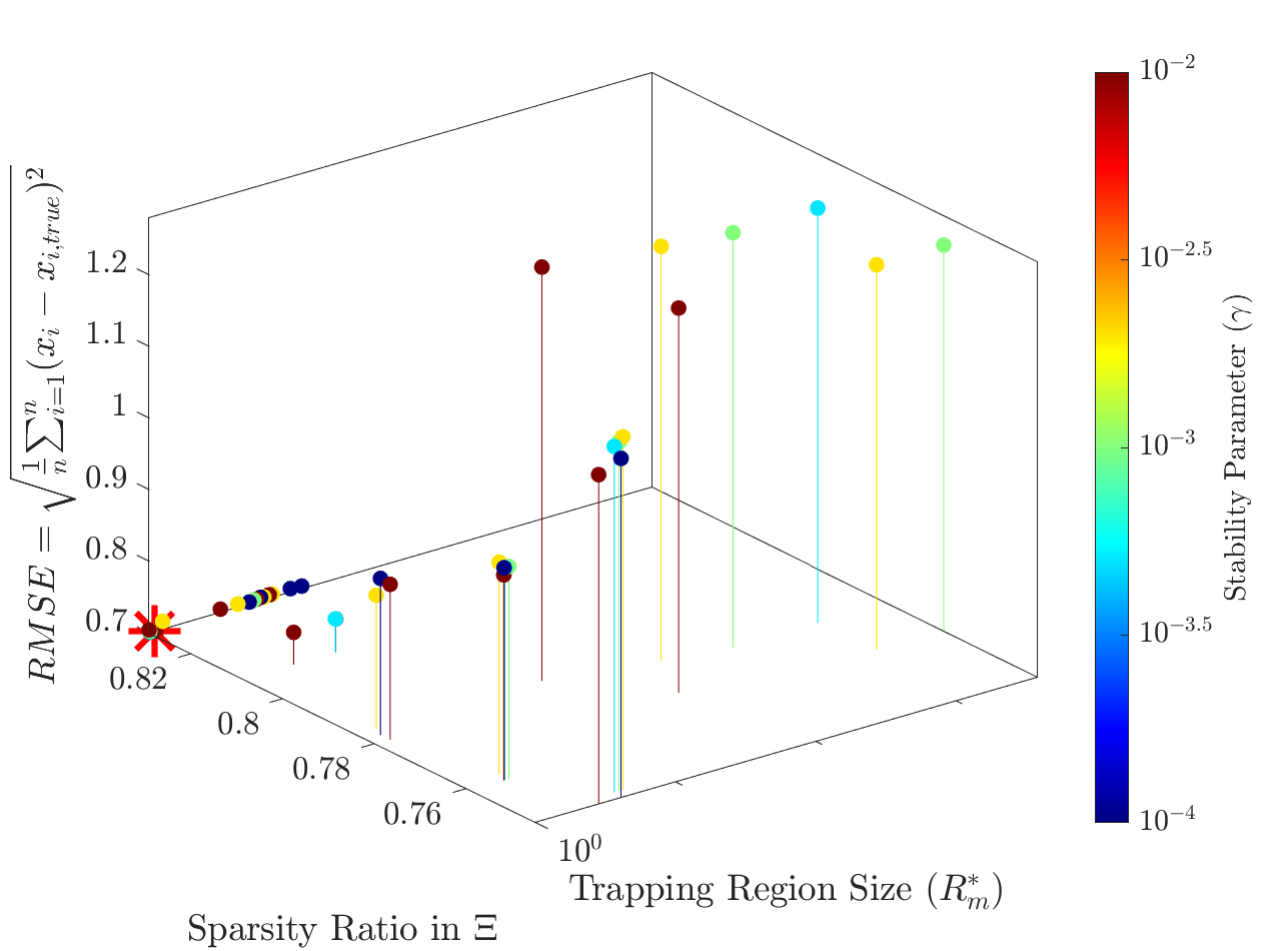}
\caption{Model selection for the nine‐state sinusoidal shear‐flow ROM. Each marker represents a candidate model obtained by sweeping the sparsity parameter $\delta$ (horizontal axis: sparsity ratio in $\Xi$) and the stability weight $\gamma$ (color scale) through our alternating regression–SDP procedure. The trapping region size $R_m^*$ (log scale, depth axis) and the root‐mean‐square error (RMSE) of the POD coefficient predictions (vertical axis) are plotted for each model. For each $\gamma$ (color), varying $\delta$ yields a vertical “stem” of points: sparser models (higher $\delta$) lie to the left, while denser models lie to the right. Notice two distinct clusters emerge, one with small trapping regions but higher RMSE (lower $\gamma$), and another with larger regions but lower RMSE (higher $\gamma$). The optimal model, marked by the red star, strikes the best compromise, achieving both low error and a tight trapping region.}
\label{fig:Error3d_9st}
\end{figure}

The model was initially trained on an arbitrary set of initial conditions and subsequently validated using a different set of initial conditions to evaluate model's ability to generalize beyond the training regime. A sweep over $\gamma$ and $\lambda$ was performed to generate a set of candidate models, and the model that best balanced a low error while maintaining a tight trapping region was selected for each case (see figure \ref{fig:Error3d_9st}). 

To determine the accuracy of the selected model, we first compared the errors between the modeled and true coefficients. The errors were computed via the Frobenius norm of the difference between the true and modeled coefficient matrices, normalized by the Frobenius norm of the true matrices. This resulted in a linear coefficient error of $\|L^* - L\|_F/\|L\|_F=0.0165 \%$ and a quadratic coefficient error of $\|Q^* - Q\|_F/\|Q\|_F=0.027\%$. The errors are all on the order of a hundredth of a percent, thus indicating that both the linear and quadratic model coefficients are accurately recovered. Next, we compared the optimal trapping region radius for our model ($R_m^*$) with that of the true system ($R_{m,true}^*$), yielding an error of $\|R_m^* - R_{m,true}^*\|_F/\|R_{m,true}^*\|_F=0.001\%$. Furthermore, the shift vector $m^*=[-0.5,0,0,0,0,0,0,0,-0.5]^\top$ of the recovered model was \emph{identical} to that of the true system. 

\begin{figure}
    \centering
    {\includegraphics[width=0.45\textwidth]{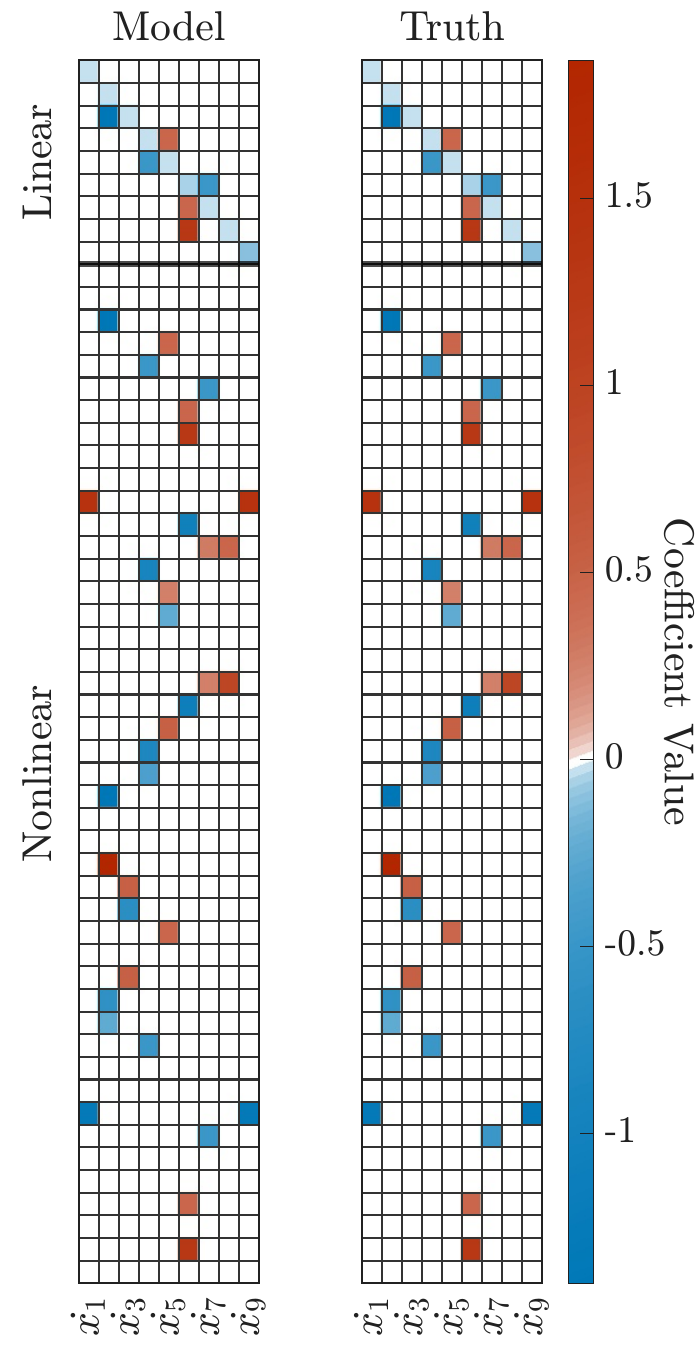}}\hfill
    \caption{Coefficient Matrices $\Xi$, for the $Re=200$ sinusoidal shear flow, with the modeled on the left and true coefficients on the right. The sparsity pattern of the modeled and true coefficients is identical. The off-diagonal Linear terms are a result of the shift $W$ to ensure that the system is centered at the origin.}
    \label{fig:Coeffs_9St}
\end{figure}

\begin{figure}
    \centering
    \subcaptionbox{Energy for $Re=200$}{\includegraphics[width=0.8\textwidth]{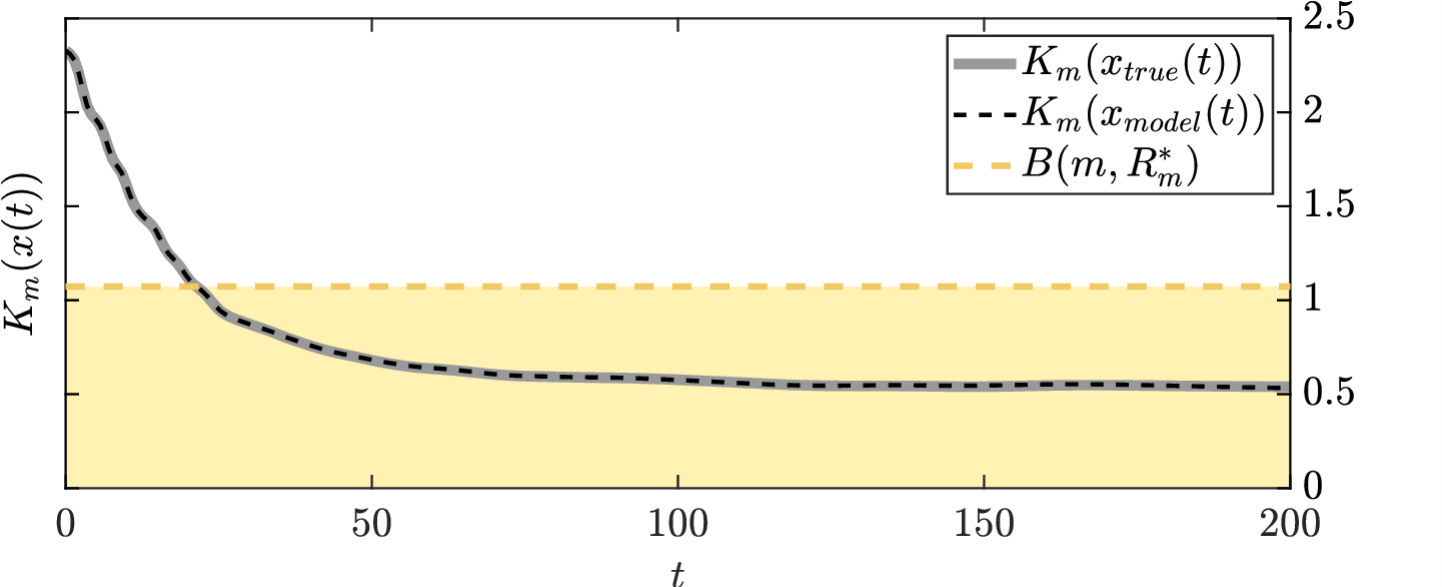}}\\
    \subcaptionbox{State Trajectories for $Re=200$}{\includegraphics[width=0.8\textwidth]{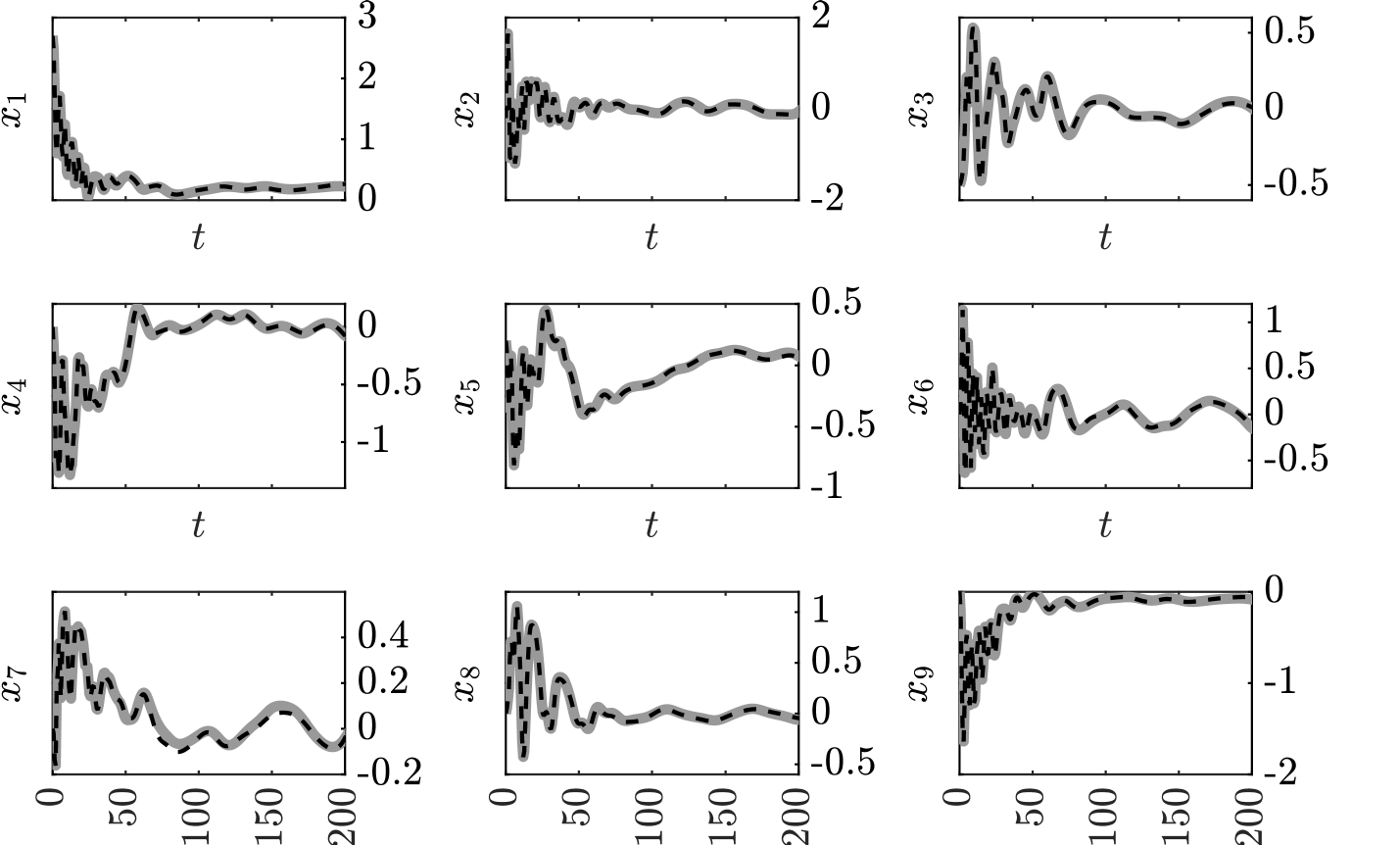}}
    \caption{The energy and state validation trajectories for $Re=200$. The gray indicates the true trajectory and the black dashed line indicates the trajectory obtained from integrating the modeled $\Xi$ from off-training initial conditions. Note that the energy decays monotonically towards the trapping region.}
    \label{fig:Energy9St_200}
\end{figure}

The Trapping-SDP modeling approach correctly recovered the sparsity pattern of the true coefficient matrices. Figure \ref{fig:Coeffs_9St} shows a comparison of the modeled coefficient matrices $\Xi^*$ and the true coefficient matrices $\Xi$, where color maps indicate active terms and inactive terms are rendered in white. The sparsity pattern of the modeled coefficients exactly matches that of the true coefficients, and the and the magnitudes match exactly.
Additionally, figure \ref{fig:Energy9St_200} shows the time evolution of the system’s energy along with the identified trapping region. The trajectories initiated from conditions outside the trapping region exhibit a monotonic decay until they enter the region, after which the energy remains bounded. The energy curves from the true model and the reconstructed model match closely.

\section{Low-Order Model of Unsteady Separation over an Airfoil}\label{sec:AirfoilResults}
\begin{figure}[h]
\centering
\includegraphics[width=0.9\textwidth]{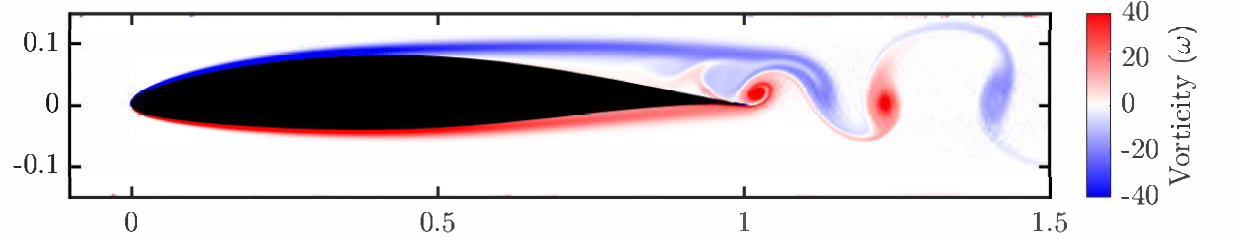}
\caption{Baseline flow over a NACA 65(1)-412 airfoil at $Re=20,000$ and $\alpha=4^\circ$.}
\label{fig:AirfoilBase}
\end{figure}

Here, we apply our framework to obtain low-order models of unsteady separation over a NACA 65(1)-412 airfoil at $Re=20,000$ (with respect to the chord) and an angle of attack of $\alpha=4^\circ$ (see figure~\ref{fig:AirfoilBase}).
Accurately modeling an airfoil’s baseline (uncontrolled) flow is critical for developing effective flow control strategies. ROMs that capture the essential flow physics of baseline flows while simplifying model complexity are especially valuable for controller design \citep{Rowley2017,Brunton2015}.
Without a faithful baseline flow model, control laws may be ineffective or even destabilizing.

\subsection{Direct Numerical Simulations and Data Curation}\label{sec:Airfoil}
For model training, we conducted direct numerical simulations~(DNS) of the baseline flow.
The flow considered is nominally incompressible; yet, the moderate Reynolds number and the strong separation bubble make it a challenging test for reduced‐order modeling: the flow exhibits a broad spectrum of energetic scales, an unsteady separation region, and a self-sustained oscillation (lift ``buzz'') whose phase must be captured accurately~\citep{Tank2021}.
We also conducted DNS of the flow response to localized pulse forcing in order to provide a rich dataset for off-training model validation.
Localized forcing was applied at six chordwise locations on the suction side of the airfoil, \[z_x/c \in \{0.1,\;0.2,\;0.3,\;0.4,\;0.45,\;0.5\},\] and at six discrete times corresponding to fractions of the natural lift‐oscillation period, \[T_{pulse} \in \{0,\,T_f/6,\,2T_f/6,\,3T_f/6,\,4T_f/6,\,5T_f/6\},\]
yielding a total of 36 validation cases.
Each pulse is an instantaneous, zero‐mass, Gaussian‐shaped forcing whose principal axis is oriented normal to the surface.
The validation cases are initialized after the pulse‐induced transient has decayed for model identification and testing of trapping‐region performance.
Specific details on the DNS are reported in appendix~\ref{sec:DNS}.

\begin{figure}
   \centering
\includegraphics[width=1\textwidth]{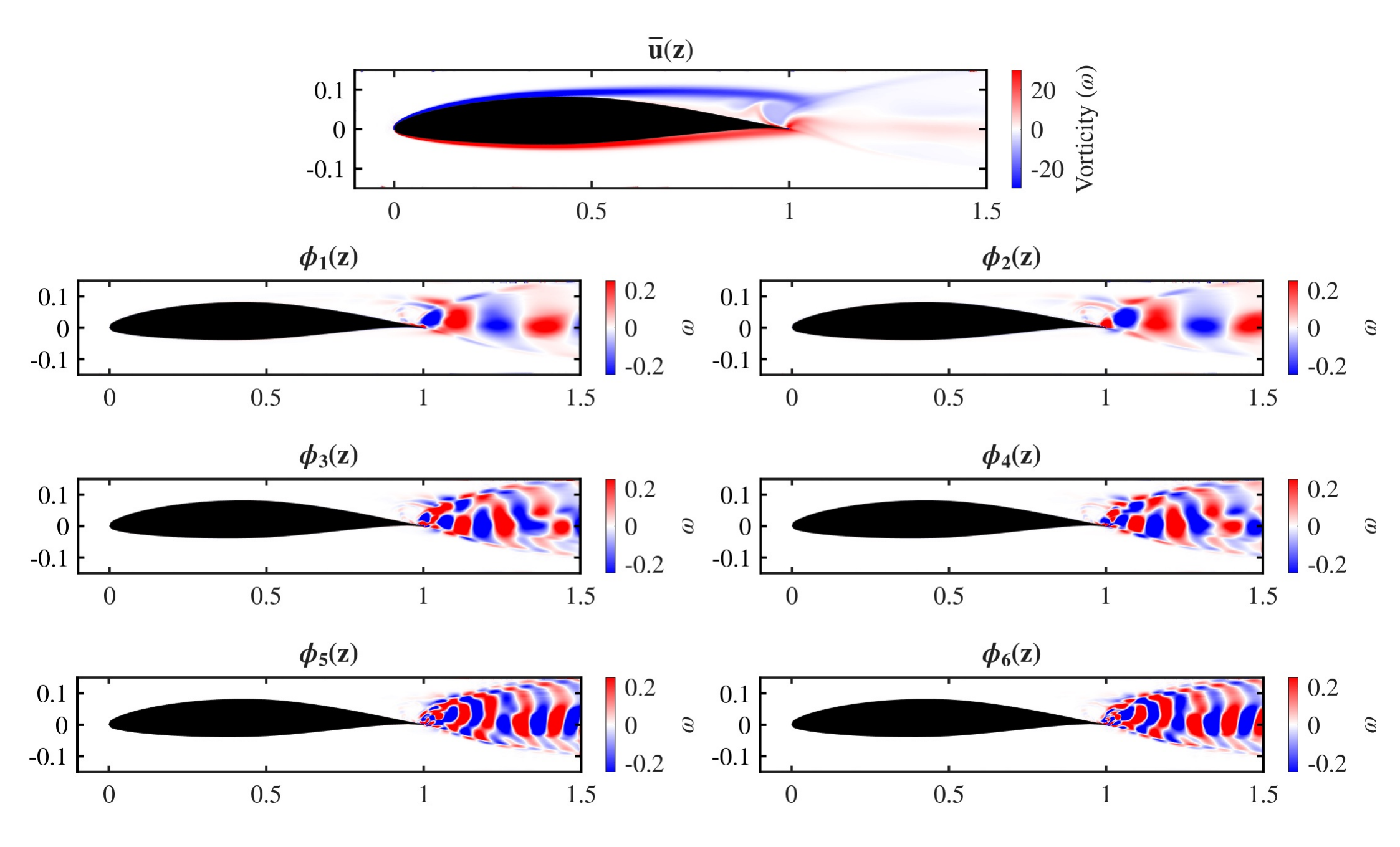}
\caption{Mean flow $\bar{\boldsymbol{u}}(\textbf{z})$ and leading six POD modes $\boldsymbol{\phi}_i(\mathbf{z})$ used in modeling flow over a NACA 65(1)-412 at $Re=20,000$ and $\alpha=4^\circ$, visualized via the associated vorticity field.}
\label{fig:PODModes}
\end{figure}

Prior to applying our modeling framework, we curate the DNS data by restricting the spatial domain to focus on the separation bubble and near-wake region ~(see figure~\ref{fig:AirfoilBase}).
The resulting spatial domain consists of $r=48,549$ spatial grid points and 550 temporal snapshots.
 Next, we decompose the velocity field $\boldsymbol{u}(\textbf{z},t)$ into a temporal-mean component $\bar{\boldsymbol{u}}(\textbf{z})$ and a fluctuating component $\boldsymbol{u'}(\textbf{z})$.
Thus,
\begin{equation} \label{POD_basis}
    \boldsymbol{u}(\textbf{z},t) = \bar{\boldsymbol{u}}(\textbf{z})+\sum_{i=1}^{n} \boldsymbol{\phi}_i(\textbf{z}) x_i(t),
\end{equation}
where we have expanded the fluctuations in a basis of proper orthogonal decomposition~(POD) modes $\boldsymbol{\phi}_i(\mathbf{z})\in\mathbb{R}^{2r}$ determined from snapshots of $\boldsymbol{u'}(\textbf{z},t)$.
In this work, we retain the leading $n=6$ POD modes (see figure~\ref{fig:PODModes}), which capture 80\% of the fluctuation kinetic energy in the baseline flow.
Thus, the state vector to be used within the modeling framework will correspond to the six POD modal coefficients $x_i$.
Note that the leading six POD modes are dominated by wake structures and that the first four modes in particular retain much of the larger-scale dynamics within the separation region above the airfoil.
The higher-order POD modes that have been excluded from the modeling exhibit finer-scale spatial structures; as such, we expect that  fine-scale features will not be fully captured in the flow reconstruction based on~\eqref{POD_basis}.

Finally, we perform a time-interpolation of the POD coefficients to refine the time-resolution ($\Delta t=0.002$) and obtain POD coefficient time-derivative information from an eighth-order central differencing scheme.
This post-processing is conducted over a shorter time-horizon than the original DNS, amounting to 14.5\% of the DNS simulation horizon.
The total number of snapshots on this finer grid available for model training is then $n_t=400$, from which snapshot data matrices $X$ and $\dot{X}$ in~\eqref{eq:SINDy} are constructed.

\subsection{Airfoil Model Results}

We apply our modeling framework to the snapshot data matrices of POD coefficients,  constructed as described above.
In doing so, we swept over the sparsity-promoting parameter~$\delta$ and the stability regularization weight~$\gamma$ to generate a family of candidate models, each constrained to admit a monotonically attracting trapping region. From this candidate pool, we selected the model that best balanced the root-mean-square (RMS) error of the modeled coefficients with respect to validation data and the size of the trapping region $B(m,R_m^*)$.

\begin{figure}
\centering
\includegraphics[width=0.8\textwidth]{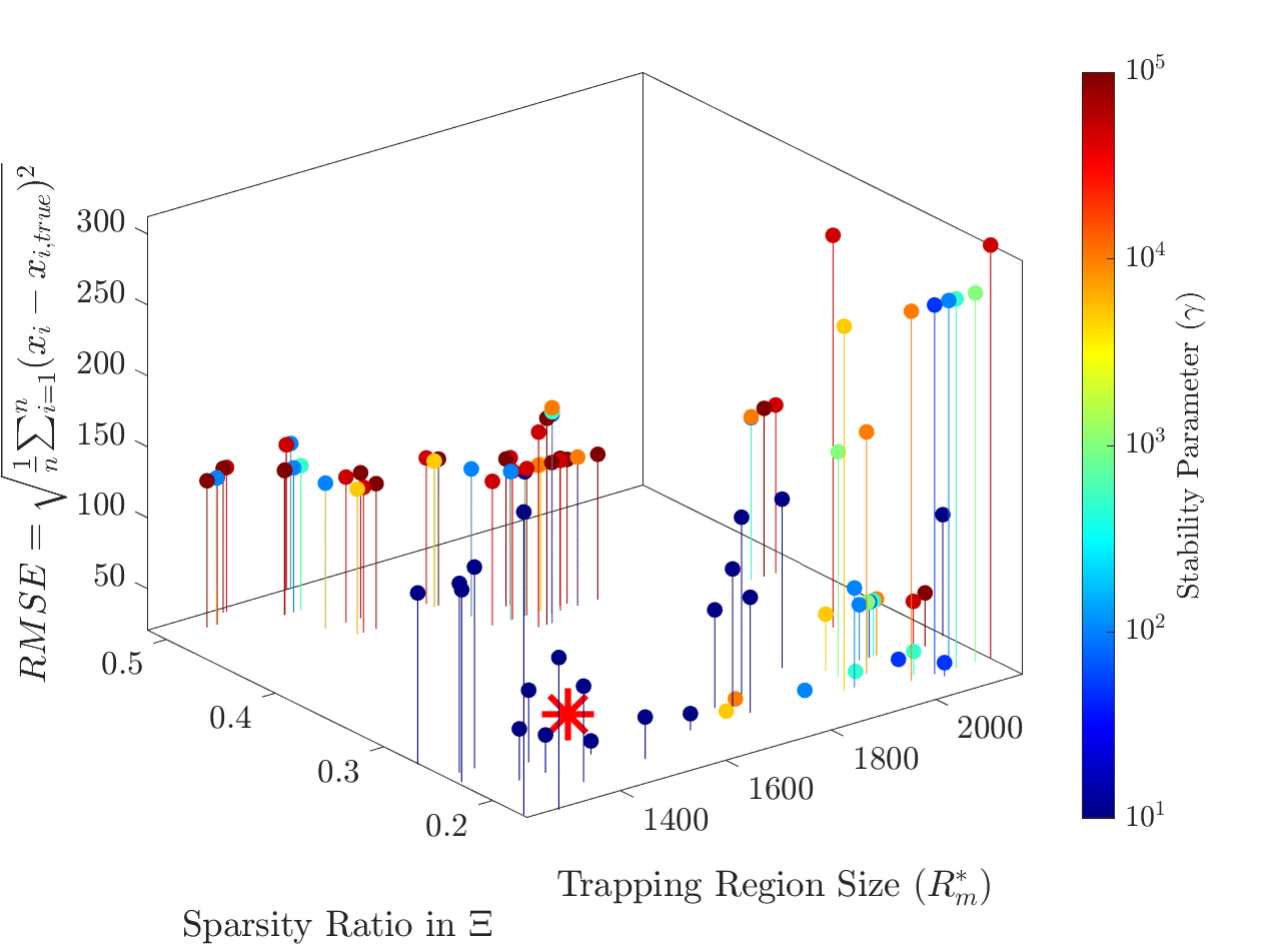}
\caption{Model selection for the six‐state airfoil ROM. Each marker represents a candidate model obtained by sweeping the sparsity parameter~$\delta$ (horizontal axis: sparsity ratio in $\Xi$) and the stability weight~$\gamma$ (color scale from dark blue ($\gamma=10^1$) to dark red ($\gamma=10^5$)). The trapping region size $R_m^*$ (depth axis) and the root-mean-square error (RMSE) of the POD coefficient predictions (vertical axis) are shown for each model. Compared to the 3‐ and 9‐state cases, the airfoil data yields a broader spread in both RMSE and trapping region size, reflecting the increased system complexity. Although many models achieve similarly low RMSE (clustered near the front of the plot), they differ substantially in their certified trapping region. We selected the red‐star model by first identifying the five models with the smallest RMSE and then choosing the one among them with the tightest bounding region. Importantly, all low‐error candidates behave well under time integration, demonstrating robustness across hyperparameter choices.}
\label{fig:Error3d}
\end{figure}

Figure~\ref{fig:Error3d} visualizes this model selection process. We plot the RMS prediction error versus sparsity and trapping region size. Models with larger trapping regions exhibit higher prediction error, as do overly sparse models. These trends suggest that an optimal trade-off exists: although sparsity is desirable, some small but important coupling terms must be retained to accurately capture the dynamics. The best-performing model was $24\%$ sparse (i.e.,~24\% of the model coefficients were set to zero) and yielded a minimal trapping region radius of $R_m^* = 1{,}457$.
Note that application of the Schlegel and Noack method to the same model yields a trapping region radius estimate of $R_m = 1{,}930$, which is roughly 32\% more conservative than the optimal estimate.
We report coefficients for the selected model in appendix~\ref{app:modelcoeffs}.

\begin{figure}
\centering
\includegraphics[width=0.3\textwidth]{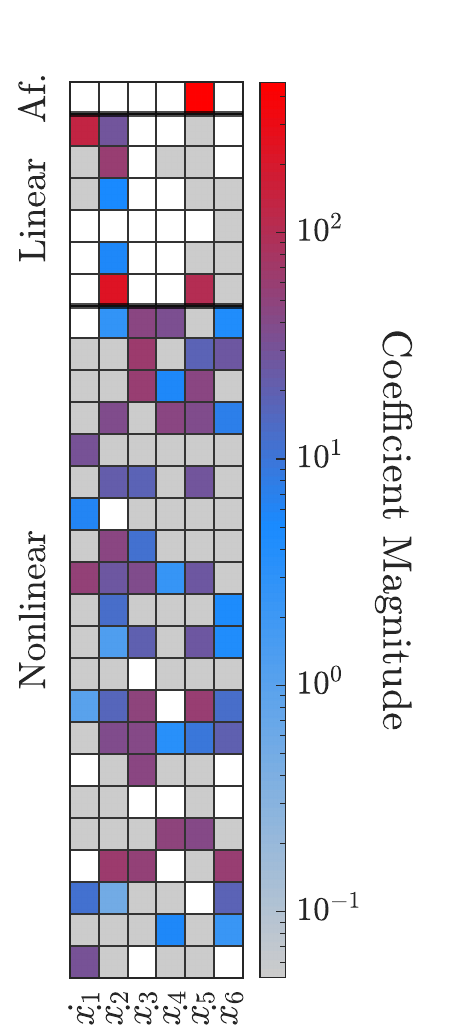}
\caption{Coefficient magnitudes of the six-state ROM for the airfoil flow. Linear, affine, and quadratic terms are shown in separate blocks; colors indicate log-scaled coefficient magnitudes, and zero (sparse) entries are rendered in white.}
\label{fig:matrix_6st}
\end{figure}

In Figure~\ref{fig:matrix_6st}, we present the learned coefficient structure of the identified six‐state model. Several features stand out. 
First, unlike our earlier models which omitted affine terms, the six‐state ROM of the airfoil benefits significantly from including affine terms in the candidate library. In practice, allowing each equation to include a constant bias term improved fit and boundedness: the learned affine shift in $\dot{x}_5$ captures a persistent mean‐flow deviation or bias that cannot be represented by purely linear or quadratic terms. This highlights the importance of incorporating affine modes when modeling flows with nonzero equilibrium offsets, ensuring the ROM can adjust its baseline state as needed for accurate long‐term predictions. 
Second, strong linear couplings are evident in the equations for $\dot{x}_1$, $\dot{x}_2$, and $\dot{x}_5$. 
In contrast, the $\dot{x}_3$ equation contains no linear or affine components at all; its evolution is governed entirely by quadratic terms, underscoring the purely nonlinear nature of that mode. Finally, mode 5 not only carries its affine shift and a strong linear coupling with mode 6, but also features significant quadratic coefficients across multiple modal pairs. This indicates that, beyond its linear and constant contributions, mode 5 is subject to rich nonlinear modal interactions. While the model is sparse, the retained coefficients (shown in gray) suggest meaningful nonlinear interactions, despite the fact that the model was trained exclusively on the unactuated baseline flow.

\begin{figure}
\centering
\includegraphics[width=0.85\textwidth]{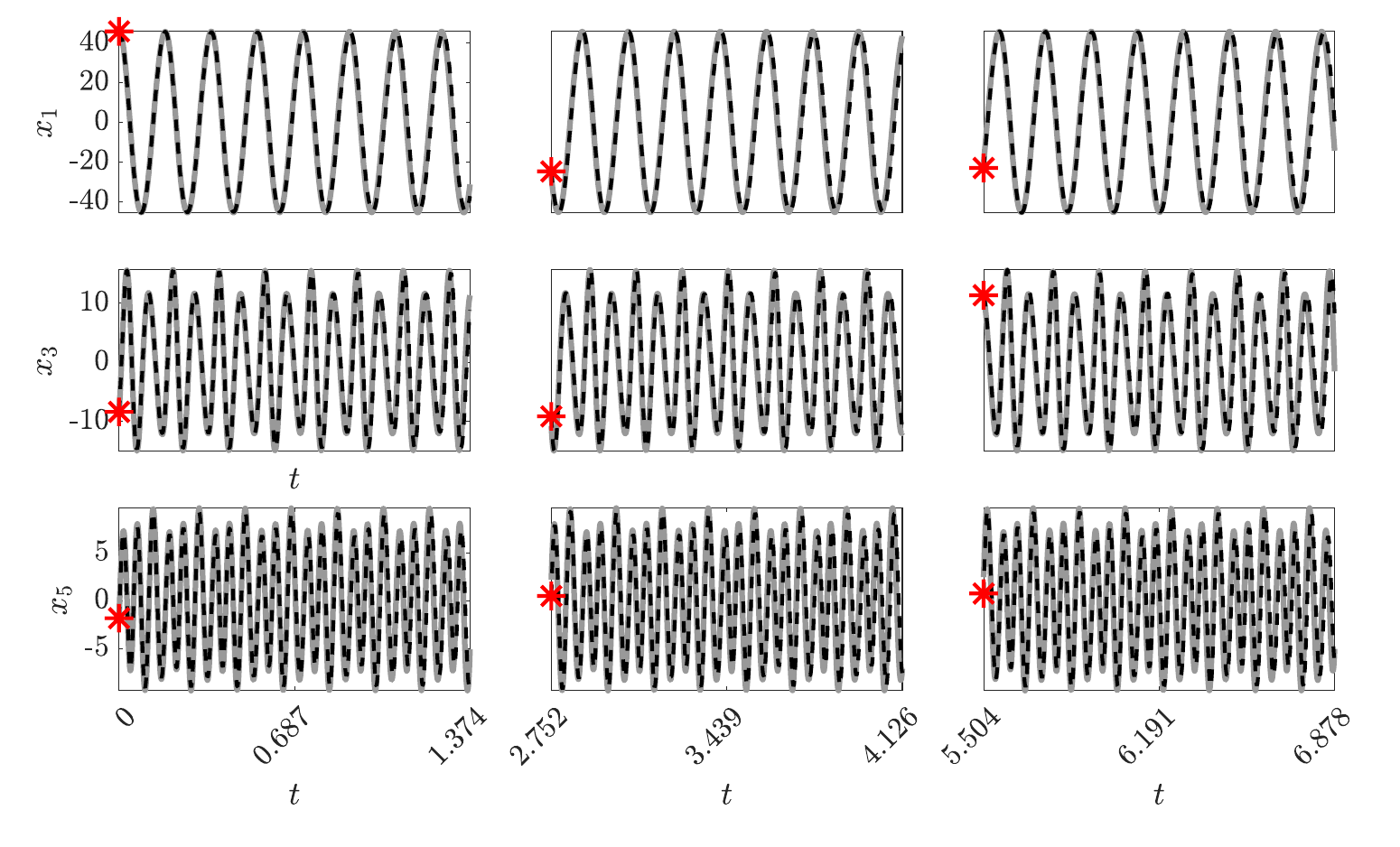}
\caption{Validation of the six-state ROM on off-training initial conditions along the baseline limit cycle. Solid gray curves show the true POD coefficients, while black dashed curves show the ROM predictions. Each column corresponds to a different phase-shifted initialization (marked by red stars). Displayed are modes $x_{1}$, $x_{3}$, and $x_{5}$; the remaining modes ($x_{2}$, $x_{4}$, $x_{6}$) exhibit similar agreement and are omitted for clarity.}
\label{fig:baselineVal}
\end{figure}

To assess generalization within the baseline limit cycle, we validated the ROM on three off-training initial conditions drawn from phase-shifted points along the same unforced trajectory. This tests whether the ROM correctly reproduces the oscillatory dynamics over long time horizons. Results for coefficients $x_1$, $x_3$, and $x_5$ are shown in Figure~\ref{fig:baselineVal}. The ROM (black dashed line) closely tracks the DNS (gray solid line) across all cases. A small phase drift is observed as time progresses, likely due to the exclusion of higher-order POD modes that may encode weak phase-coupling dynamics lost during truncation. Importantly, the model preserves the amplitude and qualitative features of the oscillation, suggesting robust internal limit cycle dynamics.

Although the model was trained only on baseline flow data, we further validated it on perturbed (pulsed) flow simulations. In these cases, the ROM was initialized with states taken \emph{after} the actuation-induced transient had largely decayed, ensuring the validation remains within the ROM’s attractor manifold. This isolates model fidelity to intrinsic dynamics rather than evaluating its response to exogenous forcing (which it was not trained to reproduce).

\begin{figure}
\centering
\includegraphics[width=0.8\textwidth]{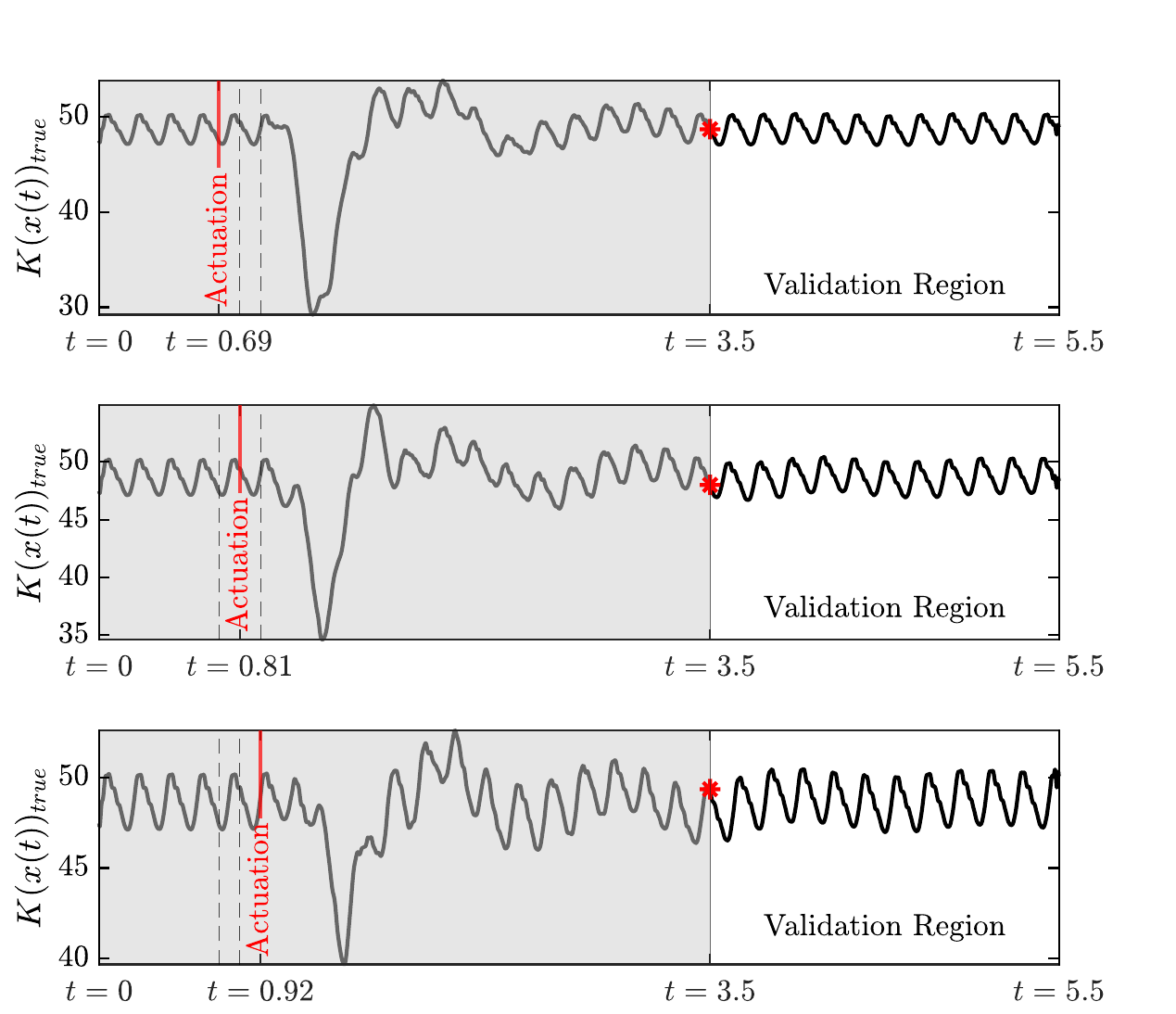}
\caption{Time series of the true kinetic energy $K(x(t))_{true}$ for pulsed flow at $x/c=0.1$, showing three sample actuation times: $t=0\,T_f/6$, $t=2\,T_f/6$, and $t=4\,T_f/6$. In each subplot, the gray-shaded interval denotes the transient phase, the solid red line marks the actuation instant, and the vertical dashed lines show the three other pulse timings for comparison. The black curve to the right of the shaded region shows the data used for validation, initialized at $t=3.5$, indicated by the red star.}
\label{fig:Ecutoff}
\end{figure}

Figure~\ref{fig:Ecutoff} shows the selected validation windows for three actuation times at $z_x/c$, with the initial state (post-transient) marked by a red star. Although some residual unsteadiness remains, this setup allows testing whether the ROM remains bounded and predictive from non-training states.

\begin{figure}
\centering
\includegraphics[width=0.8\textwidth]{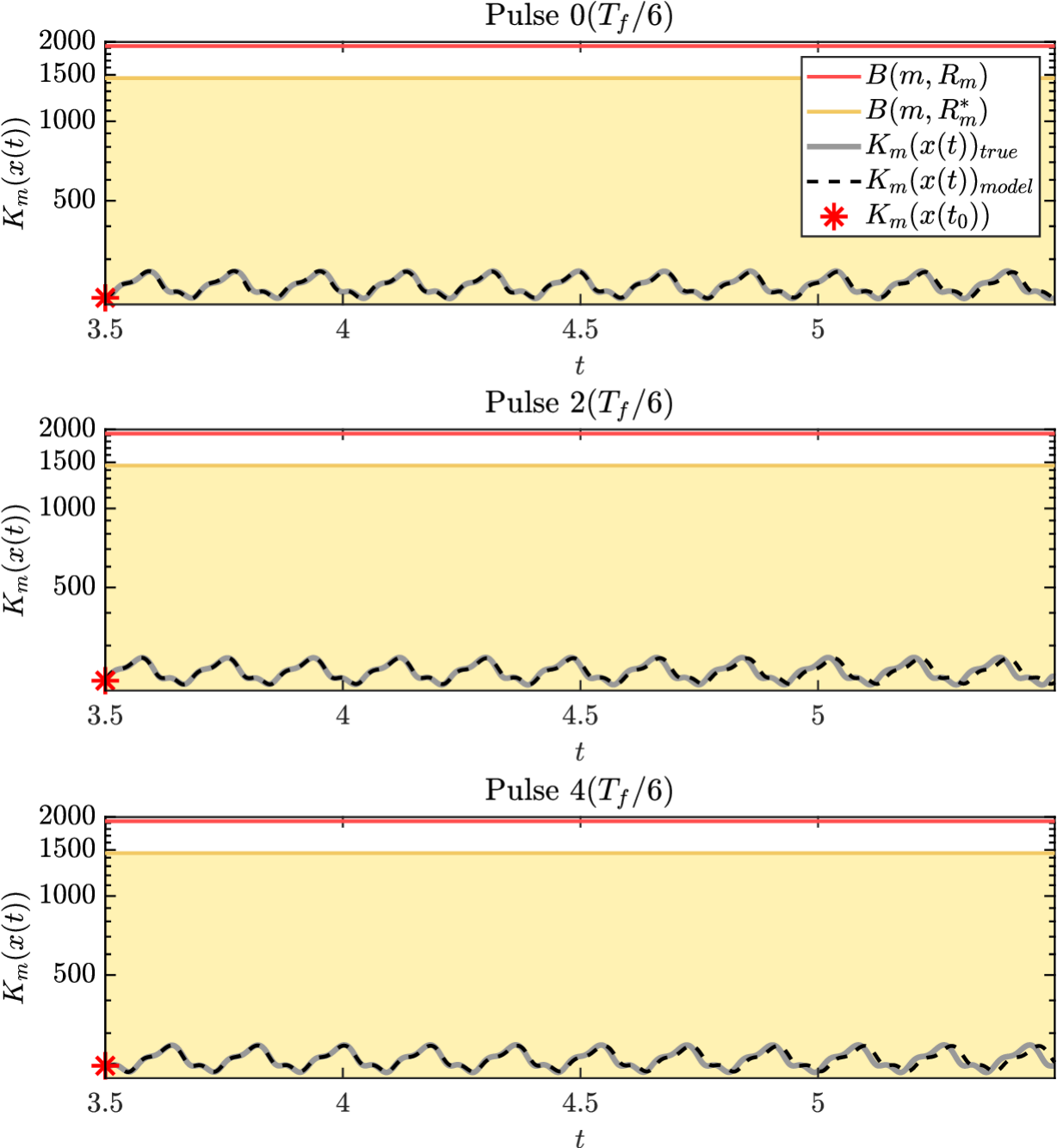}
\caption{Validation of kinetic energy evolution for pulsed flow at $x/c=0.3$ with actuation at $t=0\,T_f/6$, $2\,T_f/6$, and $4\,T_f/6$. In each panel, the solid gray line shows the true energy, the black dashed line is the ROM prediction, and the red star marks the initialization point. The shaded yellow band indicates the trapping region $B(m,R_m^*)$ computed by our SDP method, while the red horizontal line denotes the larger trapping radius obtained via the Schlegel–Noack theorem.}
\label{fig:EnergyandTR_Flow}
\end{figure}

\begin{figure}
    \centering
    \subcaptionbox{Time t=3.5}{\includegraphics[width=0.8\textwidth]{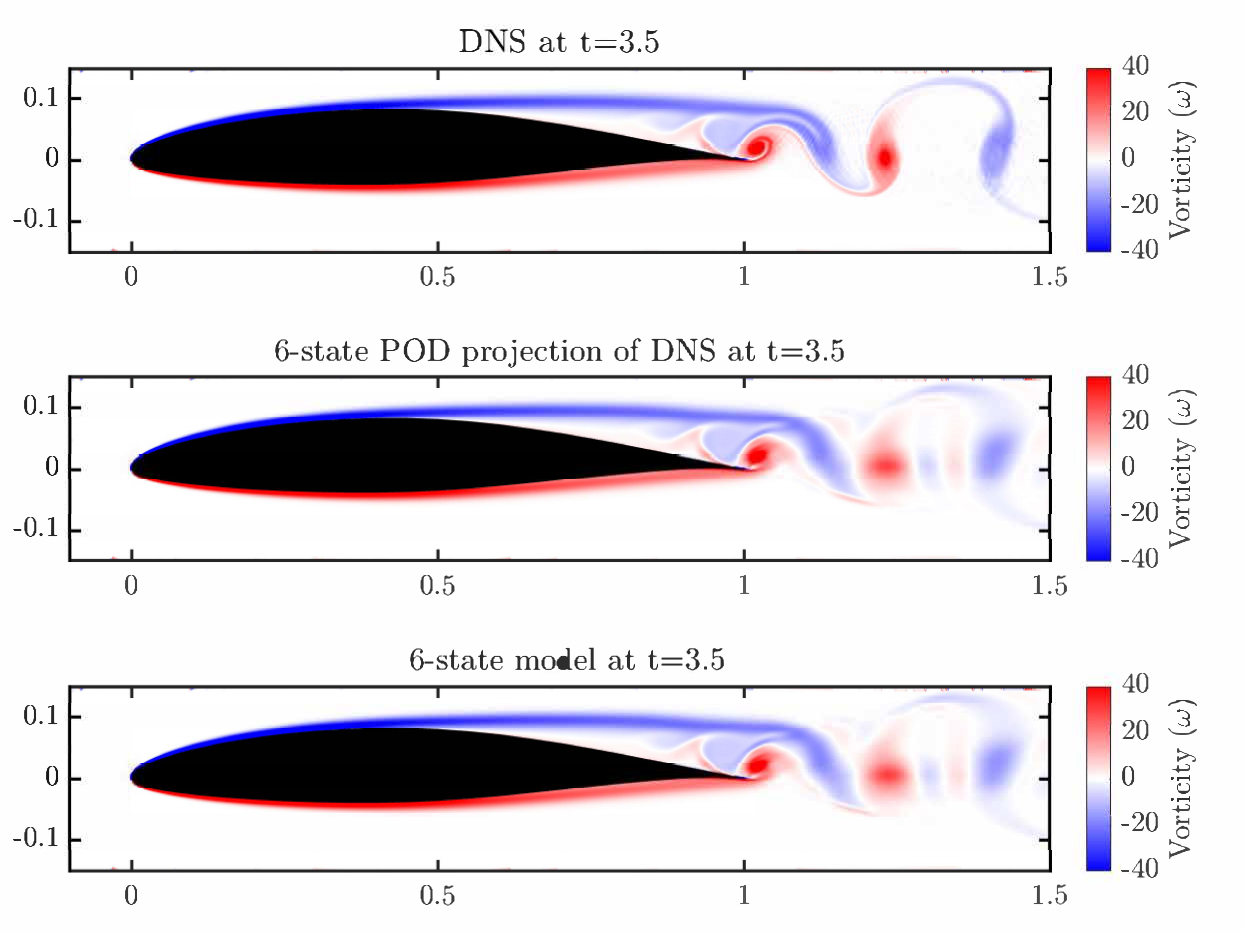}}\\
    \subcaptionbox{Time t=5.4}{\includegraphics[width=0.8\textwidth]{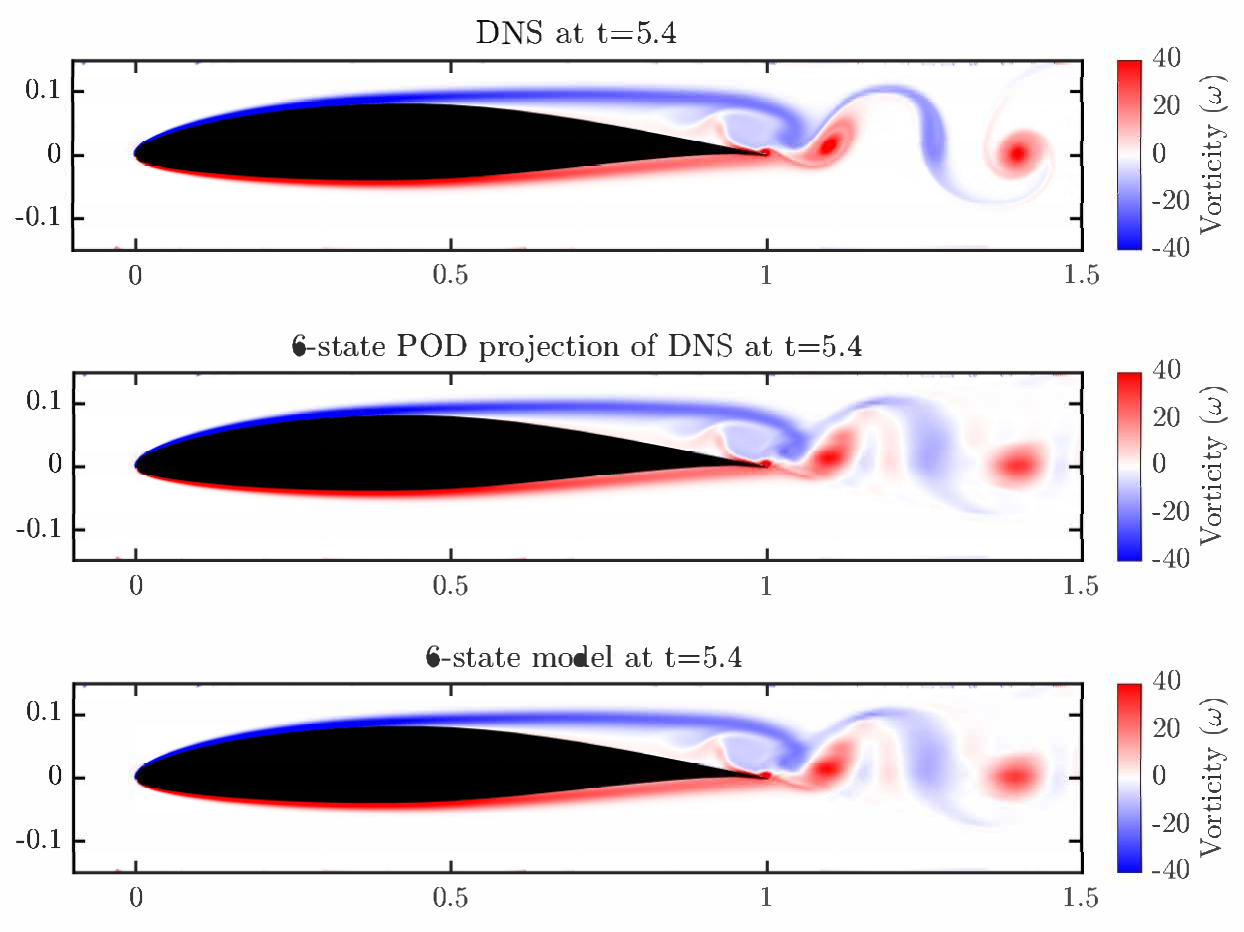}}
    \caption{Vorticity fields for the pulsed case at $x/c=0.3$ with actuation at $t=0\,T_f/6$. Shown at (a) $t=3.5$ and (b) $t=5.4$ are (top) the full DNS solution, (middle) the six-state POD projection of the DNS, and (bottom) the six-state ROM prediction.}
    \label{fig:FLOW1}
\end{figure}

\begin{figure}
    \centering
    \subcaptionbox{Time t=3.5}{\includegraphics[width=0.8\textwidth]{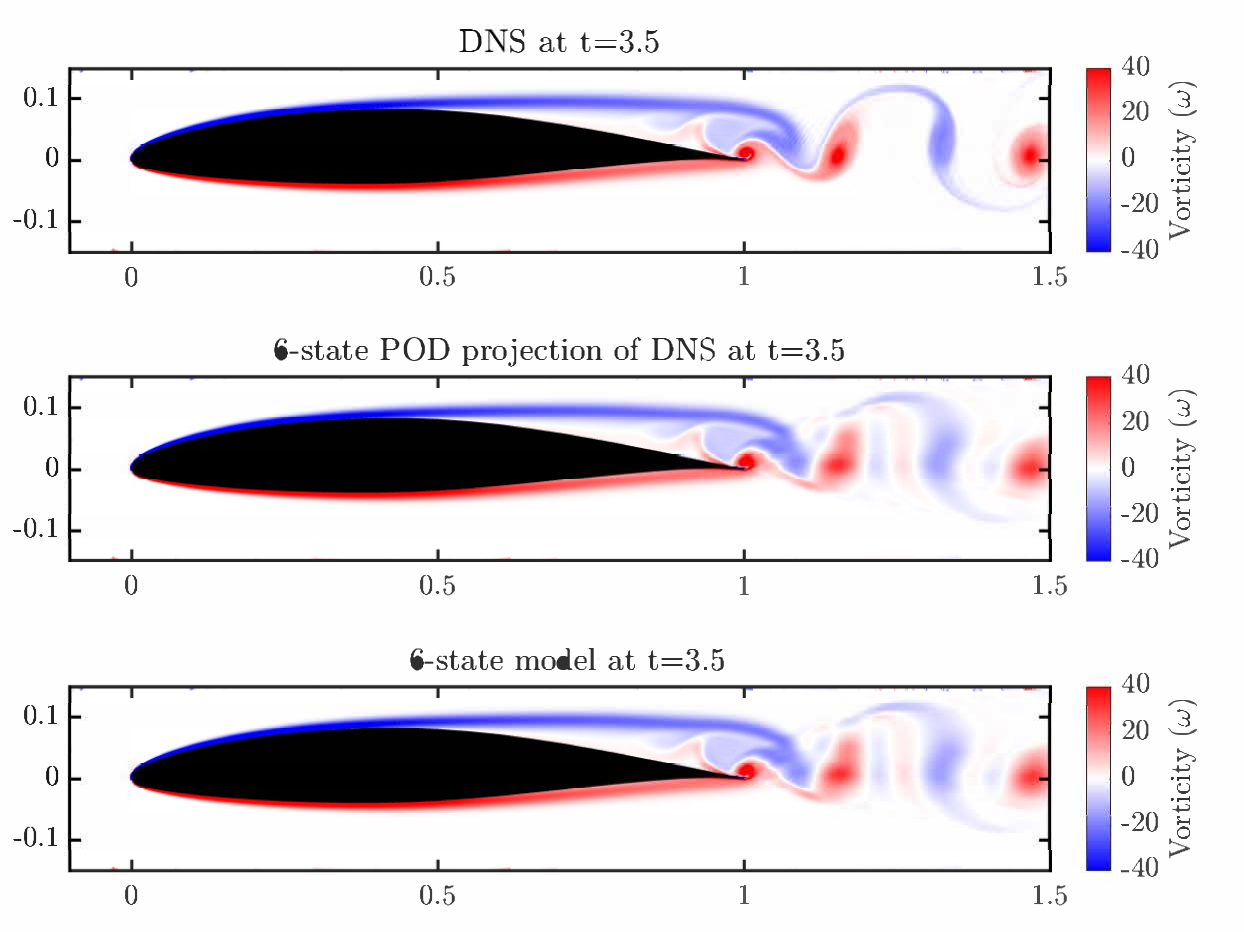}}\\
    \subcaptionbox{Time t=5.4}{\includegraphics[width=0.8\textwidth]{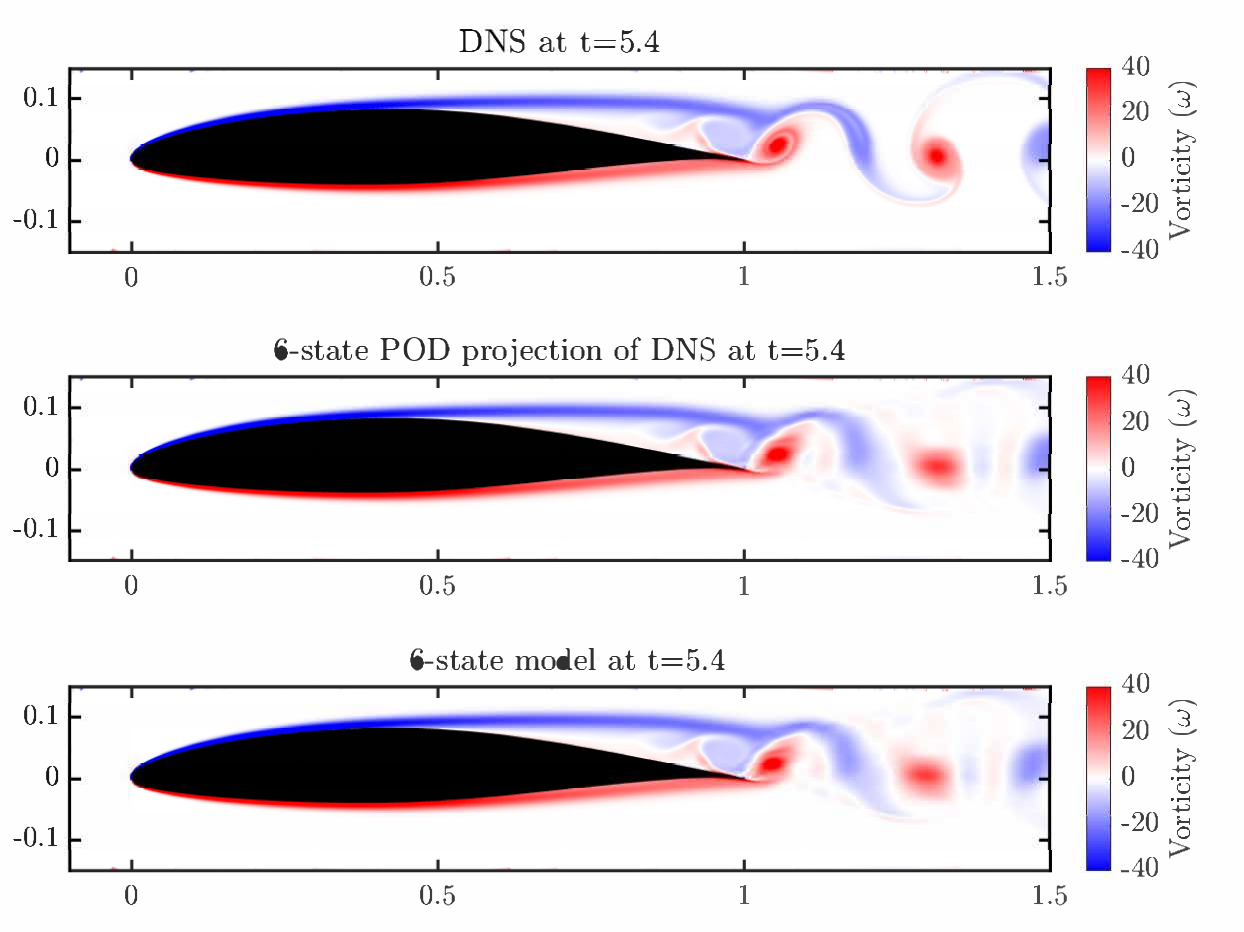}}
    \caption{Vorticity fields for the pulsed case at $x/c=0.3$ with actuation at $t=4\,T_f/6$. Shown at (a) $t=3.5$ and (b) $t=5.4$ are (top) the full DNS solution, (middle) the six-state POD projection of the DNS, and (bottom) the six-state ROM prediction.}
    \label{fig:FLOW2}
\end{figure}

Figure~\ref{fig:EnergyandTR_Flow} displays the energy evolution and trapping regions for the selected initializations. In all cases, the ROM remains safely within its trapping region for the entire simulation horizon. However, the trapping region bounds are much larger than in the previous examples. Several factors likely contribute to this comparatively large trapping region in the six‐state airfoil model. First, truncating the POD expansion to just six modes necessarily omits higher‐order interactions that would otherwise strengthen the negative‐definite component of the dynamics; without these modes, the identified quadratic operators must compensate, broadening the worst‐case energy estimate. Second, the finite POD basis itself may not perfectly orthogonalize all nonlinear couplings, introducing spurious cross‐terms that further inflate the radius required to cover every trajectory. Finally, the inclusion of a nonzero affine shift, most prominently in the $\dot x_{5}$ equation, displaces the center $m$ of the trapping ball away from the origin, which increases the minimal radius even though the actual energy peaks remain small. Nonetheless, this bound still certifies global boundedness: throughout every test case, the six‐state ROM remains inside its invariant ball and faithfully reproduces the baseline flow amplitudes. We also note that similar to the unforced case, phase drift increases with time, especially evident in the second and third scenarios. For actuation cases 1 and 5, we visualize two flow field snapshots (initial and final validation times) in Figures~\ref{fig:FLOW1} and \ref{fig:FLOW2}, comparing the ROM prediction, the DNS, and the projection of DNS onto the six POD modes. The ROM captures the flow structure well, including the separation bubble, although again, the wake exhibits higher errors due to phase shifts.

\begin{figure}[h]
\centering
\includegraphics[width=0.9\textwidth]{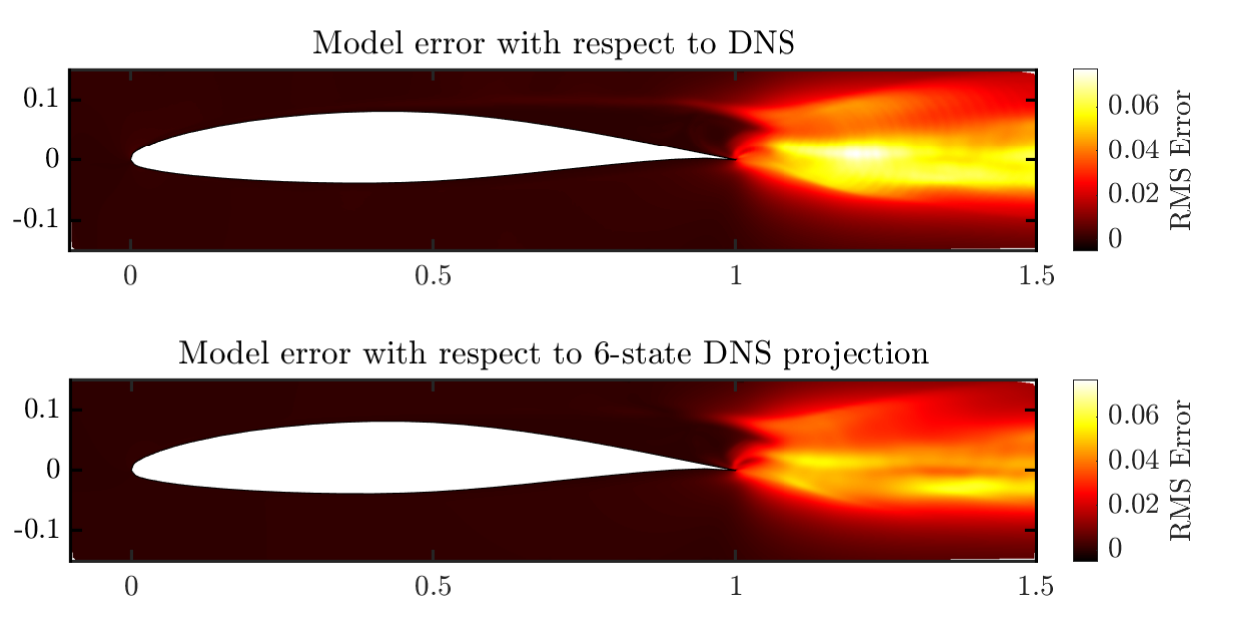}
\caption{Time-averaged RMS error of the ROM over all validation cases: \textbf{Top} compares the ROM to full DNS, and \textbf{Bottom} to the 6-state DNS projection. Error peaks in the wake downstream of the trailing edge, while the separation region above the airfoil remains accurately captured.}
\label{fig:globalFlowError}
\end{figure}

Finally, we quantified the full-field reconstruction error across all test cases in Figure~\ref{fig:globalFlowError}. The ROM-predicted velocity field was compared to both the full DNS and the six-mode projection of DNS using the relative kinetic energy of the flow. Results confirm that the ROM closely approximates the separation bubble dynamics, although error accumulates downstream in the wake due to phase drift. This is consistent with prior observations and underscores the trade-off between model simplicity and full spatial accuracy in open flows \cite{GALLETTI2007354,Callaham_Loiseau_Brunton_2023}.

\pagebreak

\section{Conclusion} \label{sec:conclusion}
In this work we have presented a novel, fully convex framework for data‐driven identification of reduced-order models (ROMs) of incompressible flows that possess certifiably optimal boundedness properties.
Our proposed Trapping-SDP modeling framework (i)~provides an unambiguous certificate of whether any lossless, globally bounded ROM exists for a given set of model parameters and hyperparameters, and (ii)~if one does exist, computes the minimal Euclidean ball into which all trajectories eventually enter and remain.  
These convex subproblems are embedded into an alternating regression and stability-enforcement loop with provable convergence properties.
We integrated our Trapping-SDP loop within the context of the SINDy modeling framework to identify models
that are simultaneously \emph{lossless}, \emph{globally bounded}, and \emph{sparse}.

We demonstrated the effectiveness of our modeling approach first on two canonical benchmark problems: the chaotic Lorenz system and a 9‐state Galerkin model of sinusoidal shear‐flow.
In these examples, our approach recovered models that faithfully reproduced the known benchmark with regards to model coefficients and minimum trapping region radius.
We then applied our approach to obtain a 6-mode model of unsteady separation over a NACA-65(1)-412 airfoil at $Re=20,000$ and $\alpha=4^\circ$ from direct numerical simulation data.
The resulting model exhibited uniformly bounded long‐time behavior under arbitrary initial conditions, while retaining only the essential affine, linear, and quadratic terms needed for accurate prediction over short- and long-time horizons.

The Trapping-SDP modeling framework was shown to yield models with optimal trapping regions in all the examples considered.
This distinguishing feature of the Trapping-SDP modeling framework facilitates the identification of models that are both physically and dynamically consistent with the governing equations, enabling predictive capabilities for both short- and long-time  horizons.
Moreover, by keeping the entire procedure convex at every step, we were able to  leverage mature off‐the‐shelf solvers to realize a tractable and reliable method for computing data-driven models with these certifiably optimal boundedness properties. We anticipate that the core idea---bringing convex, globally valid certificates into data‐driven fluids modeling---will open new pathways toward trustworthy ROMs for prediction, optimization, and control of complex flows.

\section*{Acknowledgments} This material is based upon work supported by the Air Force Office of Scientific Research under
award number FA9550-21-1-0434.

\appendix
\section{Obtaining Physics-Based Constraints}\label{apdx:triadic}

\par In order to obtain reduced-order governing equations that are consistent with the Navier--Stokes form, the physical properties of the incompressible fluid must be captured. The constrained approach described in section \ref{sec:SINDy} s active enforcement of desired conditions. To enforce physics-based constraints, the assumption that the quadratic and bi-linear terms in the Navier--Stokes equation must be energy-conserving is made.  As the state $x$ is chosen such that it is related directly to the kinetic energy of the perturbation, the constraint to make the quadratic nonlinear term energy-preserving can be written in the form: $x^\top{Q}(x){x}=0$. We note that it is also possible to use the POD-basis representation defined in \eqref{POD_basis} together with the fact that POD modes form an orthonormal set to rewrite \eqref{eq:Dynamic_System} in terms of the POD coefficients as
\begin{equation} \label{NSE_POD}
    \dot{x}_i(t)=\sum_{j=1}^{n}\sum_{k=1}^{n}Q^{(i)}_{jk}x_jx_k+\sum_{j=1}^{n}L_{ij}x_j+c_i \qquad i=1,2,...,n.
\end{equation}
The derivation of using the terms $Q_{jk}^{(i)}$, $L_{ij}$, and $c_i$ in \eqref{NSE_POD} to obtain constraints for $\Xi$ is based on triadic interactions inherent to Galerkin projection methods \citep{Rempfer_TRIAD1,Balajewicz_TRIAD2}. The approach takes advantage of the symmetric nature of the term $Q_{jk}^{(i)}$ and the fact that it retains the energy-preserving property of the non-linearity in the Navier--Stokes equations. The energy within each POD mode is expressed by $K_i(t)=\frac{1}{2}x_i^2(t)$ which is differentiated in time to obtain \citep{Rempfer_TRIAD1}
\begin{equation} \label{E_1}
    \dot{K}_i(t)=\sum_{j=1}^{n}\sum_{k=1}^{n}Q^{(i)}_{jk}x_ix_jx_k+\sum_{j=1}^{n}L_{ij}x_ix_j+c_ix_i \qquad i=1,2,...,n.
\end{equation}
Here, the constraint on $Q_{jk}^{(i)}$ means that the quadratic nonlinearity can only serve to exchange energy between modes and does not contribute to the rate of change of the total energy of the system. Using this and the fact that the sum of the individual modes constitutes the total energy of the system $K(t)=\sum^n_{i=1}x_i^2$, the following expression is written:
\begin{equation} \label{E_diff}
    \dot{K}(t)\propto\sum_{i=1}^{n}\sum_{j=1}^{n}\sum_{k=1}^{n}Q^{(i)}_{jk}x_ix_jx_k=0.
\end{equation}
Using the rules of index permutation and the symmetric property of $Q^{(i)}_{jk}$, we obtain the energy-preserving constraint 
\begin{equation} \label{CON}
Q^{(i)}_{jk}+Q^{(j)}_{ki}+Q^{(k)}_{ij}=0.
\end{equation}
Equation \eqref{CON} imposes three categories of constraints that are used to construct a constraint matrix $C$ and $d$:
\begin{enumerate}
    \item \textbf{Intrinsic constraint}: Occurs when $i=j=k$ and implies $Q^{(i)}_{ii}=0$ for any $i$. This is a result of a mode not being able to exchange energy with itself via the quadratic nonlinearity. This creates $n$ constraints, and $n$ rows are added to the matrix $C$.
    \item \textbf{Binary constraint}: Occurs when $i\ne j=k$ or $i=j\ne k$, and implies that a term of the form $\dot{x}_i\propto x_j^2$ must be balanced by a term such as $\dot{x}_j \propto x_ix_j$. Intuitively, this constraint occurs when one mode interacts quadratically with one other mode. To impose this, $n(n-1)$ rows are added to the matrix $C$ (one for each possible binary constraint).
    \item \textbf{Extrinsic constraint}: Occurs when $i\ne j\ne k$, and implies that a term $\dot{x}_i \propto x_jx_k$ involves interactions between mode $i$ and two other modes $(j,k)$. This constraint is fully triadic, and involves the energy exchange between three modes, and the energy exchange between two modes is mediated via a third mode. The complexity of these interactions result in the number of constraints added as rows to $C$ to grow rapidly: $\frac{n!}{3!(n-3)!}=\frac{1}{6}n(n-1)(n-2)$. 
\end{enumerate}
A total of $n+n(n-1)+\frac{1}{6}n(n-1)(n-2)$ rows in $C$ and elements in $d$ are therefore added upon the first solution of the constraint matrices. Constraints for small entries in the coefficient matrix $\Xi$ are then added on subsequent iterations until a solution is reached. 
This procedure results in constraints which ensure the quadratic nonlinearity conserves energy, consistent with the nonlinear physics of the incompressible Navier--Stokes equations.

\section{Governing Equations and Numerical Approximation}\label{sec:DNS}

We consider the compressible Navier-Stokes equations for conservation of mass, momentum and energy, which can be written in non-dimensional form as a system of equations where the flux vector is divided into an advective (superscript \textit{a}) and a viscous part (superscript \textit{v}):
\begin{equation}
\frac{\partial\mathbf{Q}}{\partial t} + \mathbf{F}_x^a + \mathbf{G}_y^a - \frac{1}{Re_f}\left(\mathbf{F}_x^v + \mathbf{G}_y^v\right) = \mathbf{S}.
	\label{eq:NS}
\end{equation}
where the solution vector 
$\mathbf{Q} = \left[\,\rho \quad \rho u \quad \rho v \quad \rho E\,\right]^T$
and $\rho$, $u$, $v$ and $E$ denote the density the velocity in $x$- and $y$-direction, and the total energy, respectively.
The equation of state closes the system:
$p = \rho T/\gamma M^2$.
All quantities are non-dimensionalized with respect to a characteristic, problem-dependent length
scale, reference velocity, density, and temperature yielding the non-dimensional Reynolds number,
$Re$ and Mach number, $M$.
At a sufficiently low Mach number, the flow can be modeled as incompressible to good approximation---a fact that we  exploit to obtain bounded low-order models within our data-driven modeling framework.

The system of equations is spatially approximated using a discontinuous Galerkin spectral element method (DGSEM) and integrated in time with a 4th-order explicit Runge-Kutta scheme. Gauss-Lobatto quadrature nodes are used for the spatial integration and a kinetic-energy conserving split-form approximation of the advective volume fluxes ensures stability of the scheme through cancellation of aliasing errors from the the non-linear terms.
For a detailed description of the scheme, we refer to \cite{kopriva}, \cite{gassner16} and \cite{kloseAIAA19}.

\subsection*{DNS of Pulsed Flow over Airfoil}
\begin{figure}
	\centering
	\includegraphics[width=0.60\textwidth]{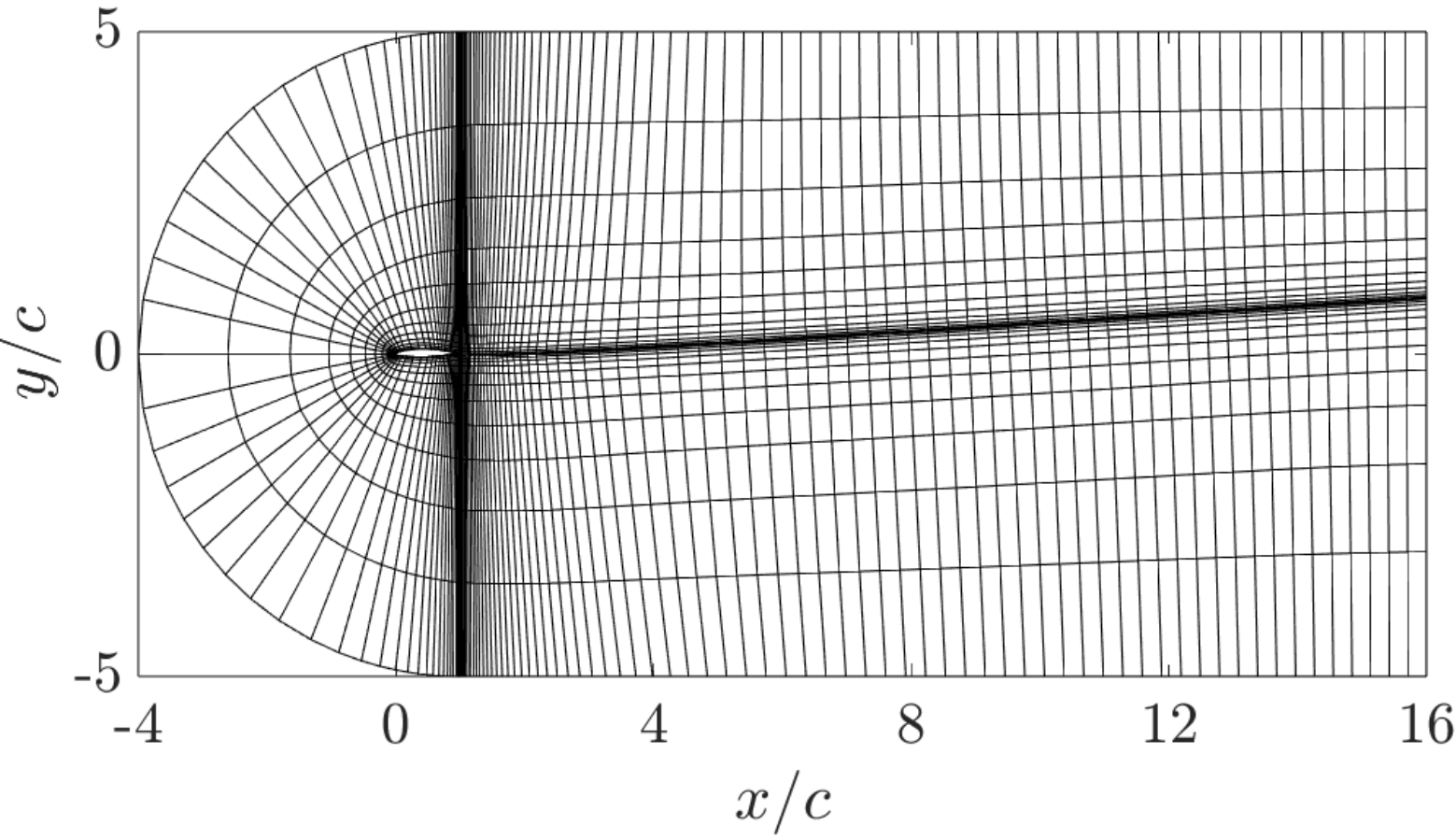}
	\caption{2D computational domain for NACA 65(1)-412. Only elements without interior Gauss-Lobatto nodes are shown.}s
	\label{fig:grid}
\end{figure}

The flow over a NACA 65(1)-412 airfoil is simulated at a Reynolds number based on the chord length of $Re = 20,000$ and a Mach number of $M = 0.3$. The low Mach number ensures a nominally incompressible flow.
The computational domain is given in Figure \ref{fig:grid} and consists of 2,256 quadrilateral elements. The domain dimensions prevent blockage and follow the recommendations in \cite{nelson16}.
The wall-boundary elements are curved and fitted to a spline representing the airfoil's surface according to \cite{nelson16}.
The boundary conditions are specified  weakly on the fluxes according to a Riemann solver \citep{JacobsOutflow}
The solution vector is approximated with a 12\textsuperscript{th} order polynomial.
giving a total of 381,264 collocation points in the domain.

The compressible Navier Stokes Equations can be written in the following way
\begin{equation}
    \frac{ \partial \boldsymbol{U}}{\partial t} + \nabla \cdot \boldsymbol{F} = \boldsymbol{S},
\end{equation}
where the solution vector $\boldsymbol{U}$ is
\begin{equation} \label{eq_var}
    \boldsymbol{U} = [ \rho \ \ \rho u \ \ \rho v \ \ \rho e ]^T,
\end{equation}
where the first entry accounts for mass conservation, second and third for momentum conservation on $x$ and $y$ respectively, and the fourth for the total energy conservation. The flux tensor is $\boldsymbol{F}$ (see \cite{nelson16} for more details on the compressible formulation).

Here we include a source term, which for a pulse, $\mathbf{S}$, is  specified  following \citep{KamphuisPulse,suzuki04}, taking the form
\begin{align} \label{universal_pulse}
    \boldsymbol{S}(t,\boldsymbol{x}) = \boldsymbol{b}(t) \odot \frac{\boldsymbol{a}}{\sqrt{(2 \pi)^n |\boldsymbol{\Sigma}^{-1}}|} \exp \left[-\frac{1}{2}(\boldsymbol{x} - \boldsymbol{c})^T \boldsymbol{\Sigma}^T (\boldsymbol{x} - \boldsymbol{c}) \right],
\end{align}
where $\odot$ denotes the Hadamard product, $\mathbf{\Sigma}$ is a $2 \times 2$ covariance matrix, $\mathbf{a}$ a vector of amplitudes, and $\mathbf{c}$ identifies the center coordinates of the Gaussian distribution. We set $\boldsymbol{\Sigma}$ to be 
\begin{equation}
    \boldsymbol{\Sigma} = \begin{bmatrix}
    \frac{\cos^2\theta}{2\sigma_x^2} + \frac{\sin^2\theta}{2\sigma_y^2} & \frac{\sin 2\theta}{4\sigma_x^2} - \frac{\sin 2\theta }{4\sigma_y^2} \\
    \frac{\sin 2\theta}{4\sigma_x^2} - \frac{\sin 2\theta }{4\sigma_y^2} & \frac{\sin^2\theta}{2\sigma_x^2} + \frac{\cos^2\theta}{2\sigma_y^2} \\
    \end{bmatrix},
\end{equation}
for $\theta \in [-45^\circ,45^\circ]$, so that the principal axes of the Gaussian Function are rotated an angle $\theta$ with respect to the $y$-axis, and with variances given by
\begin{equation}
    \sigma_x = 0.0012/\sqrt{2} \ \ \ \ ; \ \ \ \ \sigma_y = 0.0096/\sqrt{2}.
\end{equation}
The vector of amplitudes, $\mathbf{a}$, is set as
\begin{equation}
    \mathbf{a} = \left[0 \ \ , \ \  \sqrt{(2 \pi)^n |\Sigma_{mj}^{-1}|} \ \ , \ \ \sqrt{(2 \pi)^n |\Sigma_{mj}^{-1}|} \ \ , \ \ \sqrt{(2 \pi)^n |\Sigma_{mj}^{-1}|} \right],
\end{equation}
and the time-dependent component, $\mathbf{b}(t)$, as
\begin{equation}
    \mathbf{b}(t) = \delta(t-T) [ 0,\ \ -\sin(\theta), \ \ \cos(\theta), \ \ u(x,t) b_2(t) + v(x,t) b_3(t) \ ]^T,
\end{equation}
where $\delta(\tau)$ is the Dirac delta function. Note that the source is zero-mass, and the angle $\theta$ was chosen so that the principal axis of the Gaussian function is perpendicular to the airfoil surface. The Dirac delta conveniently integrates in time to the Heaviside function which means that numerically the pulsed is implemented exactly by superposing a Gaussian distribution onto the solution vector at time $T$.

The data required as input for the analysis of the optimal location and time for a pulsed control, is obtained by conducting DNS and systematically sampling  the parameter space spanned by $\mathbf{c}$ and $T$. Table (\ref{tab:test_space}) collates the values for which simulations are conducted. Here, the source center $\mathbf{c}$ in equation (\ref{universal_pulse}) is located at a distance of $2\sigma_y$ into the flow, measured perpendicular to the airfoil suction side, at the listed cord-length locations (following \cite{suzuki04}), and times are presented as sixth's of the natural period for the lift. Combinations of all locations and activation times for the source give us a total of 36 test points.

\begin{table}
\begin{center} 
\begin{tabular}{||c c c c c c c||}
 \hline
 $x/c$ & $0.1$ & $0.2$ & $0.3$ & $0.4$ & $0.45$ & $0.6$ \\ [0.5ex] 
 \hline
 $T$ & $0(T_f/6)$ & $1(T_f/6)$ & $2(T_f/6)$ & $3(T_f/6)$ & $4(T_f/6)$ & $5(T_f/6)$ \\ [0.5ex] 
 \hline
\end{tabular}
\end{center}
\caption{Locations, $x/c$, on the suction side of the airfoil for the source actuation, and respective times, $T$, as fractions of base flow natural period $T_f$.}
\label{tab:test_space}
\end{table}

\section{Coefficient Data for the 6-Mode Model of Unsteady Separation}\label{app:modelcoeffs}
This appendix reports coefficients for the affine, linear, and quadratic terms associated with the 6-mode model of unsteady separation over a NACA 65(1)-412 airfoil at $Re=20,000$ (with respect to the chord) and an angle of attack of $\alpha=4^\circ$ identified and discussed in Section~\ref{sec:Airfoil}. 
Note the reported values are not specified to full floating point precision here.

\begin{equation*}
c=
\begin{bmatrix}
      0 & 0 & 0 & 0 & 461.8307 & 0\\ 
    
\end{bmatrix}^\top
\end{equation*}

\begin{equation*}
L=
\begin{bmatrix}
      1.1462 & 27.9268 & 0 & 0 & -1.12 & 0\\ 
      -27.9622 & 0.47724 & 0 & -0.75324 & -1.3887 & 0\\ 
      -9.7073 & 4.9442 & 0 & 0 & -12.9241 & -6.3824\\ 
      0 & 0 & 0 & 0 & 0 & -8.66\\
      0 & 5.6722 & 0 & 0 & -32.2873 & -88.2074\\
      0 & 1.9627 & 0 & 0 & 98.4096 & -20.0578\\
\end{bmatrix}
\end{equation*}

\begin{equation*}
Q^{(1)}=
\begin{bmatrix}
 0 & -0.010943 & -0.18916 & -0.13865 & 0.13571 & -0.018203\\ 
 -0.010943 & 0.049632 & -0.14343 & 0.22095 & -0.017403 & -0.20099\\ 
 -0.18916 & -0.14343 & -0.5033 & 0.0042717 & -0.12765 & 0\\ 
 -0.13865 & 0.22095 & 0.0042717 & -0.37788 & -0.10491 & 0\\ 
 0.13571 & -0.017403 & -0.12765 & -0.10491 & 0.098613 & -0.0034017\\ 
 -0.018203 & -0.20099 & 0 & 0 & -0.0034017 & 0.25128\\
\end{bmatrix}
\end{equation*}

\begin{equation*}
Q^{(2)}=
\begin{bmatrix}
0.021886 & -0.024816 & -0.12021 & 0.15575 & -0.058758 & 0.09508\\ -0.024816 & 0 & 0.19448 & 0.11146 & 0.05388 & 0.005476\\ 
-0.12021 & 0.19448 & -0.093743 & 0.066773 & 0.158 & -0.015765\\ 
0.15575 & 0.11146 & 0.066773 & -0.020351 & -0.06229 & 0.26202\\ 
-0.058758 & 0.05388 & 0.158 & -0.06229 & 0.0042669 & -0.12548\\ 
0.09508 & 0.005476 & -0.015765 & 0.26202 & -0.12548 & -0.035263\\
\end{bmatrix}
\end{equation*}

\begin{equation*}
Q^{(3)}=
\begin{bmatrix}
0.37832 & 0.26364 & 0.25165 & -0.027281 & -0.064825 & 0.077515\\ 
0.26364 & -0.38895 & 0.046872 & 0.1518 & -0.076879 & 0.083674\\ 
0.25165 & 0.046872 & 0 & 0.2017 & 0.16747 & 0.18692\\ 
-0.027281 & 0.1518 & 0.2017 & 0 & -0.25651 & 0.23072\\ 
-0.064825 & -0.076879 & 0.16747 & -0.25651 & -0.077262 & -0.013341\\ 
0.077515 & 0.083674 & 0.18692 & 0.23072 & -0.013341 & 0\\
\end{bmatrix}
\end{equation*}

\begin{equation*}
Q^{(4)}=
\begin{bmatrix}
0.27731 & -0.3767 & 0.023009 & 0.18894 & -0.058475 & -0.031335\\ 
-0.3767 & -0.22293 & -0.21857 & 0.010175 & -0.050336 & -0.087282\\ 
0.023009 & -0.21857 & -0.4034 & 0 & 0.01456 & -0.28309\\ 
0.18894 & 0.010175 & 0 & 0 & 0.19856 & 0\\ 
-0.058475 & -0.050336 & 0.01456 & 0.19856 & -0.34793 & 0.022695\\ 
-0.031335 & -0.087282 & -0.28309 & 0 & 0.022695 & -0.49437\\
\end{bmatrix}
\end{equation*}

\begin{equation*}
Q^{(5)}=
\begin{bmatrix}
-0.27142 & 0.076162 & 0.19248 & 0.16339 & -0.049307 & 0.1252\\ 
0.076162 & -0.10776 & -0.081122 & 0.11263 & -0.0021334 & 0.10619\\ 
0.19248 & -0.081122 & -0.33495 & 0.24195 & 0.038631 & -0.069319\\ 
0.16339 & 0.11263 & 0.24195 & -0.39712 & 0.17397 & -0.017783\\ 
-0.049307 & -0.0021334 & 0.038631 & 0.17397 & 0 & -0.077685\\ 
0.1252 & 0.10619 & -0.069319 & -0.017783 & -0.077685 & -0.018803\\
\end{bmatrix}
\end{equation*}

\begin{equation*}
Q^{(6)}=
\begin{bmatrix}
0.036407 & 0.10591 & -0.077515 & 0.031335 & -0.1218 & -0.12564\\ 
0.10591 & -0.010952 & -0.06791 & -0.17473 & 0.019291 & 0.017632\\ 
-0.077515 & -0.06791 & -0.37384 & 0.052372 & 0.082661 & 0\\ 
0.031335 & -0.17473 & 0.052372 & 0 & -0.0049113 & 0.24718\\ 
-0.1218 & 0.019291 & 0.082661 & -0.0049113 & 0.15537 & 0.0094017\\ 
-0.12564 & 0.017632 & 0 & 0.24718 & 0.0094017 & 0\\
\end{bmatrix}
\end{equation*}

 \bibliographystyle{ieeetr}
 \bibliography{refs}

\end{document}